\documentclass[aps,prb,twocolumn,floatfix,superscriptaddress]{revtex4}

\usepackage{graphicx}
\usepackage{amsmath}
\usepackage{amsfonts}
\usepackage{amssymb}

\begin{document}
\bibliographystyle{unsrt}

\newcommand{\norm}[1]{\ensuremath{| #1 |}}
\newcommand{\aver}[1]{\ensuremath{\langle #1 \rangle}}
\newcommand{\ket}[1]{\ensuremath{| #1 \rangle}}
\newcommand{\fref}[1]{Fig.~\ref{#1}}
\newcommand{\sref}[1]{Sec.~\ref{#1}}
\newcommand{\aref}[1]{Appendix~\ref{#1}}
\newcommand{\hc}{\text{H.c.}}

\newcommand{\figwidth}{0.8\columnwidth}

\title{Orbital current patterns in doped two-leg $Cu$-$O$ Hubbard ladders}

\author{P. Chudzinski}
\affiliation{Laboratoire de Physique des Solides, Bat. 510, Universit\'e Paris-Sud 11, Centre d'Orsay, 91405 Orsay Cedex, France}
\author{M. Gabay}
\affiliation{Laboratoire de Physique des Solides, Bat. 510, Universit\'e Paris-Sud 11, Centre d'Orsay, 91405 Orsay Cedex, France}
\author{T. Giamarchi}
\affiliation{DPMC-MaNEP, University of Geneva, 24 Quai Ernest-Ansermet CH-1211 Geneva, Switzerland }

\begin{abstract}
In the weak coupling limit, we investigate two-leg ladders with a unit cell containing both $Cu$ and $O$ atoms, as a function of doping. For purely repulsive interactions, using bosonization, we find significant differences with the single orbital case: a completely massless quantum critical regime is obtained for a finite range of carrier concentration. In a broad region of the phase diagram the ground state consists of a pattern of orbital currents plus a density wave.  NMR properties of the $Cu$ and $O$ nuclei are presented for the various phases. 
\end{abstract}
\maketitle

\section{Introduction}

Over the past two decades, the description of strongly
correlated electron materials has been one of the most actively
pursued problems in condensed matter physics. When the strength of
Coulomb interactions between carriers is of the order of (or
larger than) their kinetic energy, many new remarkable phenomena
may occur. Their fingerprints are seen in experiments done on systems such as
cuprate compounds with high temperature superconductivity\cite{PALeeetal_review} , cobaltites
with large termopower\cite{Herbert_large_thermo_Co}, magnesium oxides with colossal
magnetoresistance\cite{Salamon_CMR}, or heavy fermions\cite{Stewart_HF}.
Among these materials, cuprates play a special role. At half filling they are insulators
with antiferromagnetic (AF) order, but, with doping, a sequence of phases is observed
including spin-glass, pseudogap, d-type superconductivity (SCd) and eventually
Fermi-like behavior for very large carrier concentrations.

Unfortunately there is, to date, no consensus on a theoretical model that
would allow one to describe the physics of the $Cu$-$O$ planes. In order to 
get insight into this strong correlation problem, the study of ladder structures \cite{dagotto_ladder_review,nagata_ladder_single}
has proven quite useful. Ladders are the simplest systems that interpolate between one- and two-
dimension.  They constitute the quasi- one dimensional analog of the $Cu$-$O$ sheets and, because of the reduced dimensionality, even weak interactions lead to dramatic effects. In the one dimensional (1D) case, the weak- and strong- interaction limits are usually smoothly connected \cite{giamarchi_book_1d}. Controlled non-perturbative methods -- like
bosonization or conformal field theory -- and numerical techniques can be used to analyse these systems. 
%

Compounds characterized by a ladder structure \cite{dagotto_ladder_review,nagata_ladder_single}, such as SrCuO, have been synthesized. They show a variety of unusual
properties, for example large magnetic fluctuations, SCd with purely
repulsive interactions and metal-insulator transitions under high
pressure \cite{piskunov_ladder_nmr, piskunov04_sr14cu24o41_nmr,fujiwara03_ladder_supra, imai_NMR_doped_2ladder,kumagai_NMR_2ladder}. For these materials, increasing the pressure
amounts to changing the bandwidth, and hence the ratio of Coulomb to
kinetic energies in the ladder structure.

These experimental developments provided a strong incentive for theorists
to study two-leg ladders with Hubbard interactions between electrons.
In the weakly interacting limit, renormalization group (RG) analysis
\cite{solyom_revue_1d} was used to
explore their phase diagram\cite{varma_2bands,fabrizio_2ch_rg,schulz_moriond,kuroki_2chain_phases,ledermann_2band_spinless,balents_2ch,lin_Nchains}. Tsuchiizu and
Suzumura \cite{tsuchiizu_2leg_NMR}, and Tsuchiizu and Furusaki
\cite{tsuchiizu_2leg_firstorder} performed an RG analysis in bosonization language
in order to explore the regime of dopings close to half filling.
Using current algebra, where spin rotational symmetry was introduced \emph{a priori} in order
to derive RG equations, Balents and
Fisher \cite{balents_2ch,lin_Nchains} established the phase diagram of two
leg ladder versus doping, showing that there was
interesting physics at finite dopings. They identified a sequence of phases,
labelled $CnSm$
with $n$ (m) gapless charge (spin) modes.
 Numerical DMRG calculations focused on the large U limit
\cite{poilblanc_2ch_mc, noack_dmrg_2ch, srinivasan_DMRG}, the so
called t-J approximation at half filling \cite{nagaosa_2ch,scalapino_t-Jcorrelation, roux_CDW+current}. The relevance of interchain hoppings on the low energy physics
was also adressed \cite{khveshenko_2chain,tsuchiizu_confinement_ladder_long, tsuchiizu_confinement_spinful,Yoshioka_2chains-phases}.

The above mentioned papers all assume that, in the low energy limit,
the $Cu$-$O$ system can be reduced to an effective single orbital model. In the
context of two dimensional (2D) cuprate materials, such  reduction to a single orbital model 
was proposed by Zhang and Rice\cite{zhang_rice_singlet} and it allowed one to derive phase diagrams
for these systems \cite{Allouletal_RMP,PALeeetal_review}. However this simplification was called into question, and it 
was pointed out that it is necessary to retain the full three band nature of the model in order to capture 
the important physics \cite{varma_3band_model, CVarma_orbitalcurrents}. This issue becomes particularly relevant
when one examines the possible existence of orbital current phases. Such phases were initially proposed for the
Hubbard model \cite{affleck_marston}. They were subsequently analyzed by various authors
\cite{kotliar_liu_dwave_slavebosons, lederer_superconductivity_flux_ref, zhang_flux_ref, PALeeetal_review},
but in slave boson and in numerical calculations one finds that they are unstable. For single band ladder models,
controlled calculations appropriate to 1D reveal that for special choices of interactions -- which must include
non local terms -- staggered flux patterns are stable. This phase breaks the translational symmetry of the lattice
\cite{orignac_2chain_long, schollwock_CDW+current}. According to some authors
\cite{chakravarty_ddw_pseudogap, PALeeetal_review}, the 2D version of this state (the DDW phase) describes 
the pseudogap phase of the cuprates. An alternative type of orbital current pattern, which preserves the lattice
translational symmetry, was advocated to describe the pseudogap phase 
\cite{varma_3band_model, CVarma_orbitalcurrents}. It then requires using a three band model. Recent experimental
data taken from neutron measurements\cite{BFauque_neutrons} and polar
Kerr effect\cite{Xia_optOAF} would be consistent with the latter proposal, but more studies are clearly needed
to fully corroborate this scenario.


Motivated by these considerations, 
Lee, Marston and Fjarestad \cite{lee_marston_CuO}
(see also Ref.~\onlinecite{jeckelmann_DMRG})
generalized the system of RG equations written in current
algebra language by Balents and Fisher to study the $Cu$-$O$ Hubbard ladder.
Their work was
however limited to the half-filled case, where umklapp terms dominate the
physics, giving rise to Mott transitions. In a recent rapid communication\cite{chudzinski_ladder_rapid}
we outlined the method which allowed us to map out the full diagram of the $Cu$-$O$ ladder
as a function of doping.

The aim of the present paper is to provide details of our derivation, and to present new results
which are experimentally testable.
In our work,
oxygen atoms are taken into account at each
calculation step, which allows us to probe their influence. First, 
 they lead to new types of phases compared with the single orbital case:
a Luttinger liquid (LL) regime is found for a fine range of dopings and, in a broad 
region of the phase diagram, the ground state displays an orbital current pattern plus density
 wave quasi long-range order.
Our study thus underscores the importance of including these additional degrees of freedom in the structure, in
particular with regards to the existence and to the stability of currents patterns.
Although our results 
have been derived for the specific case of ladders, they have 
potential relevance to the physics of 2D cuprate materials as well. 
Second, spectroscopic tools measuring local properties, such as NMR,
are predicted to give different signatures depending on wether they probe Cu
or O sites. In the large U limit, for 2D cuprates, it is
believed that spin fluctuations on oxygen sites merely track those
on the copper sites\cite{Allouletal_RMP}. The advantage of revisiting the
issue in a
quasi-1D context is that one can monitor spin excitations on
oxygen atoms both in the small and large U limits using
bosonization techniques. This is done in the present paper for
various dopings in the small U limit; we do find differences
between the NMR signal on the copper and oxygen atoms at low
temperature, when gaps set in, but not at higher temperature in the
Luttinger liquid (LL) regime.

The paper is organized as follows: in \sref{sec:model} we define the
model including the interactions relevant to the low energy
physics. In the continuum limit the quadratic part of the
Hamiltonian is diagonal in a particular basis, $B_{o}$. We give the
relations between this basis, the bonding/antibonding basis
$B_{o/\pi}$ (relevant in the non-interacting case) and the
total/transverse density basis $B_{+-}$ (the most appropriate to
write "backward" interactions).

In section III we present a new method which allows one to set up
the RG equations in the case of generic doping. \cite{chudzinski_ladder_rapid}. One of its salient features 
is that it treats the
rotation of $B_{o}$ with respect to $B_{+-}$ during the flow. This effect
needs to be taken into account in order to perform all
calculations properly. We list the resulting set of equations;
their derivation is presented in the Appendix.

The various flows and the resulting phase diagram are given in
section IV. First we assess the impact of the additional degrees of freedom, hence
we set all Coulomb interactions pertaining to the $O$ atoms and direct interoxygen
hoppings to zero.
Some
of the results obtained in previous work\cite{balents_2ch} can
now be checked using our improved RG method. In constrast with the single orbital case, we find an
 intermediate doping range where all spin and gap modes are massless (i.e a quantum critical line).
Next,
 interactions involving oxygen atoms and
hoppings between these atoms are introduced. We find that interoxygen hoppings promote
a phase of orbital currents and we analyze its structure.
Spin-rotational symmetry was not
imposed a priori, but we checked that the required property was
preserved during the flow. This provides a check on the
consistency of our calculations. In the case of massive regimes
the evolution of the gaps with doping is shown. We briefly examine
the impact of umklapp terms which are present at half filling.

In section V, we compute spin correlation functions, which allows us
to derive the Knight shifts K and the relaxation rates $T_{1}$ for Cu
and O nuclei. There are several improved features in our work. In
Ref.~\onlinecite{tsuchiizu_2leg_NMR}, spin-spin correlation functions
were calculated in the low temperature limit, using Majorana
pseudo-fermions for the spin part; this assumes that gap
opening in the spin and in the charge modes occur at well separated $T$.
In bosonization language, spin-spin
correlation function are easily obtained in all cases. For instance, if one treats
spin and charge density fluctuations on equal footing, one shows
that the uniform part of the susceptibility approaches a quantum
critical point as doping increases and that, at low temperatures in the gapped phase, the staggered part 
gives a different temperature dependence for each atom in the elementary cell.
We also discuss physical implications of the orbital current phase.


\section{The model} \label{sec:model}
\subsection{Hubbard Hamiltonian for $Cu-O$ two-leg ladders}

We consider a two-leg
ladder with a unit cell containing two $Cu$ and five $O$
atoms. Two edge oxygen sites are included because they would provide connections with
neighboring ladders (which are not considered in the present work).
\begin{figure}
\centering
  \includegraphics[width=.4\textwidth, height=7cm]{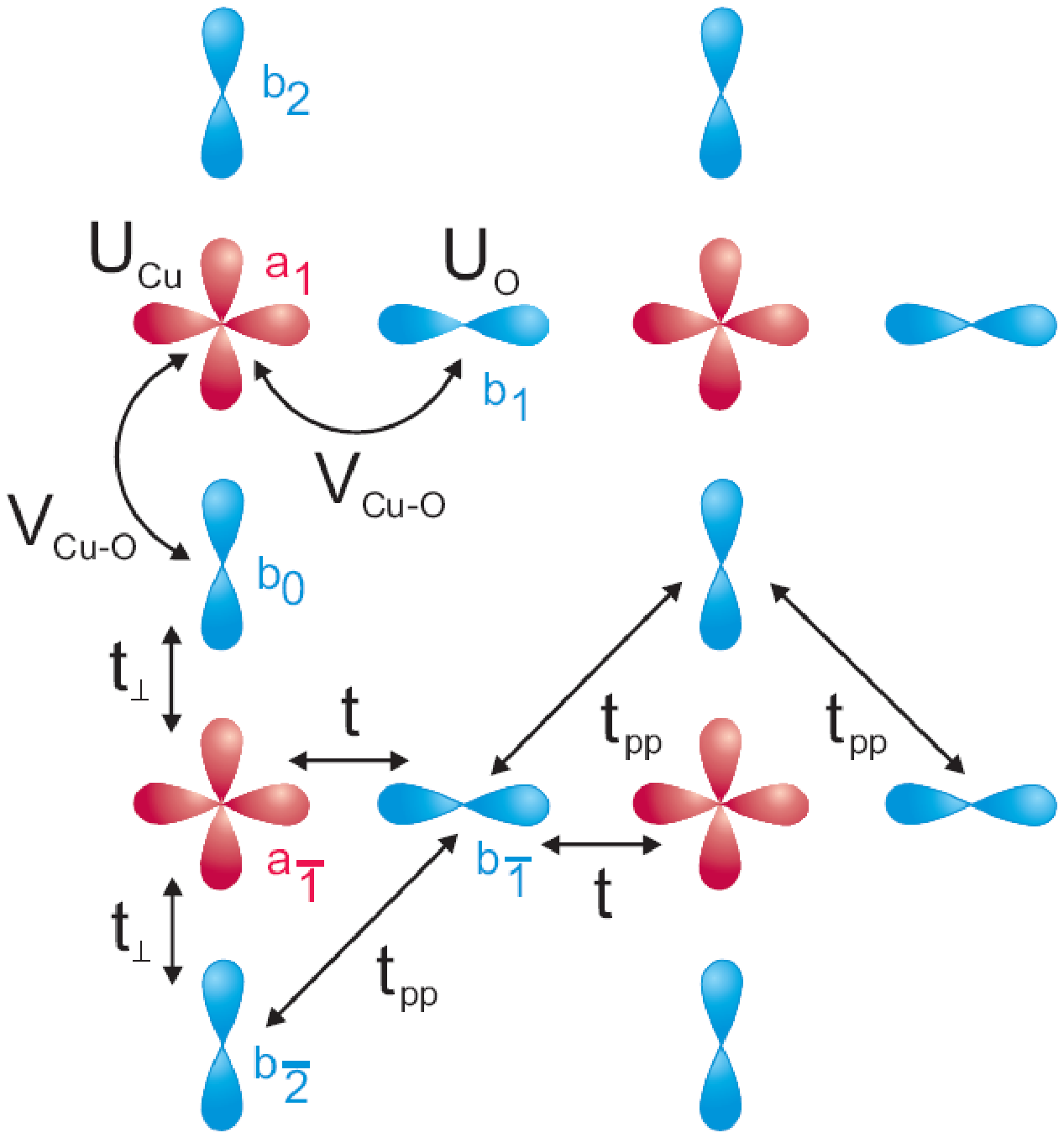}
  \caption{
 Energy contributions to the hamiltonian of the $Cu$-$O$ Hubbard ladder; the figure shows two-unit cells.
 Subscripts in the $a$, $b$ annihilation operators track the coordinate
 of the various atoms in a cell.
 }\label{fig:structure}
\end{figure}
The hamiltonian of this system is divided in two parts: the
kinetic energy of electrons moving on the lattice $H_{T}$ and
electron interactions $H_{int}$
\begin{equation}\label{hubb}
    H=H_{T}+H_{int}
\end{equation}
The explicit form of the first, tight-binding, part is
\begin{widetext}
\begin{multline}\label{hubbT}
    H_{T}=\sum_{j\sigma}( \sum_{m\in Cu}\epsilon_{Cu}
    n_{mj\sigma}+ \sum_{m\in O}\epsilon_{O}
    n_{mj\sigma}- \sum_{m\in Cu} t
    [a_{mj\sigma}^{\dag}(b_{mj\sigma}+b_{mj-1,\sigma})+\hc] \\
- \sum_{m\in Cu} t_{\bot}
    [a_{mj\sigma}^{\dag}(b_{m+1,j\sigma}+b_{m-1,j\sigma})+\hc])
-\sum_{m=\in O(leg)} t_{pp}
    [b_{mj\sigma}^{\dag}(b_{m+1,j\sigma}+b_{m-1,j\sigma}+b_{m+1,j-1\sigma}+b_{m-1,j-1\sigma})+\hc])
\end{multline}
\end{widetext}
where $a_{mj\sigma}(b_{mj\sigma})$  annihilates holes with spin $\sigma$ on a copper (oxygen) site,
$j$ labels cells along the chain and $m$ labels the atoms within
each cell. $n_{mj\sigma}=a_{mj\sigma}^{\dag} a_{mj\sigma}$ is the density of particles on site $m$, and
we use here hole notation such that $t$, $t_{\bot}$, $t_{pp}$ are all positive.
$\epsilon=\epsilon_{O}-\epsilon_{Cu}$ is the difference between
the oxygen and copper on-site energies.

 LDA determined values \cite{Muller-Rice_LDA}  of
the parameters pertaining to \emph{SrCuO} systems show that interladder hopping amplitudes
are at least \emph{one order} of magnitude smaller than their
intraladder counterparts, so that the two-leg ladder description is an excellent starting point for these
compounds. Inside the elementary cell, $t$ and $t_{\bot}$ are the dominant hoppings and
their values are comparable. The difference between the electronic 
$Cu$- $d$ and $O$- $p$ state energies $E_{Cu}$ and $E_{O}$, is about \emph{0.5t}. 
There does not appear to be
\emph{ab initio} determinations
of Coulomb terms for \emph{SrCuO} ladders, but from what is
known for cuprates, we may estimate a local $U$
of order $5t$ for the $Cu$ sites, meaning a strongly interacting regime.
In the following we will use constant values of the band parameters
$t=t_{\perp}=1$ and $\epsilon=0.5$, and treat the other observables ($U_{Cu}$, $t_{pp}/t$, $U_{O}/U_{Cu}$,
 $V_{Cu-O}/U_{Cu}$) as tunable variables.
 In order to gain insight into the physics of the multiband case, we analyze the above model using a
renormalization group procedure in the interactions, i.e we assume that all of these are smaller than the kinetic 
energy; hence, the validity of the solutions cannot be ascertained, in the event when some of the interactions
were to grow so large during the flow that they became of the order of the bandwidth.
As was stated above, the experimental regime corresponds to a situation where Coulomb terms are sizable.
Nevertheless the RG approach allows one to obtain a full analytical solution of this complicated problem, and to 
make  detailed comparisons with the physics of the one band system. Furthermore, for the case of the single band ladder 
one finds that the physical properties in the weakly- and strongly- interacting limits are smoothly connected.
We will come back to that point when we discuss our results.

Eigenvalues and eigenvectors of the non-interacting part are
simply obtained by Fourier transforming $H_{T}$. Since $\epsilon$ is of order $t$, we
neglect the non-bonding and antibonding higher energy bands which
are mostly of $p$-type character, and this reduces the model to
two lowest lying bands crossing the Fermi energy. The Hamiltonian
is:
\begin{equation}\label{HTdiag}
    H_{T}=\sum_{k\alpha\sigma}e_{\alpha}(k)n_{k\alpha\sigma}
\end{equation}
where $\alpha=0,\pi$ denotes the bands and $\sigma$ are spin indices. The operators corresponding to the eigenstates of $H_T$ are 
\begin{equation}\label{bazy1}
    a_{mk\sigma}\;\; (\textrm{or}\;\; b_{mk\sigma}) = \sum_{\alpha} \lambda_{m\alpha} a_{\alpha k\sigma}
\end{equation}
$e_{\alpha}(k)$  are the eigenvalues of $H_{T}$  (the $Cu$-$O$ distance is set to unity), and $\lambda_{m\alpha}$
are the amplitudes of the overlaps of the eigenvectors with the atomic wavefunctions in the unit cell. 
This defines the bonding ($o$) and antibonding ($\pi$) eigenbasis $B_{o/\pi}$. 
For $t_{pp}\neq 0$, the $o$ and  $\pi$ energy bands are the two lowest, real solutions of
the characteristic equation
\begin{widetext}
\begin{equation}
\begin{split}
(\epsilon-e_{o}(k))(3
t_{\perp}^{2}+\epsilon^{2}-e_{o}(k)^{2})-2(1+\cos(k))(-6 t_{pp}
t_{\perp} t-t^{2}(\epsilon-e_{o}(k))+3
t_{pp}^{2}(\epsilon+e_{o}(k)))=0 \\
(\epsilon-e_{\pi}(k))(
t_{\perp}^{2}+\epsilon^{2}-e_{\pi}(k)^{2})-2(1+\cos(k))(-2 t_{pp}
t_{\perp} t -t^{2}(\epsilon-e_{\pi}(k))+
t_{pp}^{2}(\epsilon+e_{\pi}(k)))=0 \\
e_{n}=\epsilon
\end{split}
\end{equation}
\end{widetext}
Including $t_{pp}$ increases the values of the
$\lambda_{bi\alpha}$ for the $O$ atoms and makes the $o$ and
$\pi$ bands more asymmetric, but there are still only two bands
crossing the Fermi energy, so that the analysis remains valid. We note, however, that the 
contribution of the oxygen $p$-orbital
perpendicular to the one participating in the $Cu$-$O$ bonding increases as $t_{pp}$
grows larger, until, for $t_{pp}>0.5t$,
it dominates that of the copper d-orbital. Hence, we confine the range of variation of $t_{pp}$
 to  $0$-$0.5t$.

The interaction part in fermionic language is given by:
\begin{widetext}
\begin{equation}\label{hubbInt}
    H_{int}=\sum_{j} (\sum_{m\in Cu} U_{Cu}n_{mj\uparrow}n_{mj\downarrow} + \sum_{m\in O}U_{O}
    n_{mj\uparrow}n_{mj\downarrow}+ \sum_{m\in Cu, n\in O}\sum_{\sigma, \sigma'}V_{Cu-O}n_{mj\sigma}n_{nj\sigma'})
\end{equation}
\end{widetext}

\subsection{The continuum limit and bosonization}

We now express the Hamiltonian in bosonic representation. The procedure is 
standard \cite{giamarchi_book_1d, voit_bosonization_revue} and we only outline the main steps here.
We linearize the dispersion relation in the vicinity of the Fermi energy:
\begin{equation}\label{contin}
    H_{T}=\sum_{|q|<Q} \sum_{r\alpha\sigma}r q V_{F\alpha} a^{\dag}_{\alpha r
    q\sigma} a_{\alpha r q\sigma}
\end{equation}
$r=\pm 1$ denotes right and left movers, with momenta close to their respective $\pm k_F$, $Q$ is a momentum cutoff.
The boson phase fields denoted by $\phi_{\sigma\alpha}(x)$ are
introduced for each fermion specie. $\sigma\alpha$ contains spin and band indices, \emph{x} is the spatial
coordinate along the ladder. Fermionic operators are expressed in term of the bosonic field $\phi_{\sigma\alpha}(x)$
and $\theta_{\sigma\alpha}(x)$ related to carriers fluctuations, by
\begin{equation} \label{eq:mapbos}
 \psi_{r\sigma\alpha}\simeq\eta_{r\sigma\alpha}\exp(\imath k_{F\alpha})\exp(\imath (r\phi_{\sigma\alpha}-\theta_{\sigma\alpha}))
\end{equation}
where $\eta_{\sigma\alpha}$ are the Klein factors which satisfy the
required anticommutation relations for fermions. These
$\eta_{\sigma\alpha}$ do not contain any spatial dependence and they commute
with the Hamiltonian operator. They only influence the form of
the order operator in bosonic language (through terms of the form $\eta_{\sigma\alpha}\eta_{\sigma\alpha'}$)
and the signs of the non-linear couplings through a $\Gamma$
coefficient (the eigenvalue of the
$\eta_{\sigma\alpha}\eta_{\sigma\alpha'}\eta_{\sigma\alpha''}\eta_{\sigma\alpha'''}$ operator).
The  operator is unitary so $\Gamma^{2}=1$. This equality applies also to linear 
combinations of fields (change of basis); 
 Following Ref.~\onlinecite{tsuchiizu_2leg_firstorder}, we choose 
$\Gamma=+1$ in the $\sigma+/-$ basis (see below).
We also introduce the phase field $\theta_{\sigma\alpha}(x)$; its
spatial derivative $\Pi_{\sigma\alpha}(x)=\partial_x \theta_{\sigma\alpha}(x)$ is canonically conjugated to
$\phi_{\sigma\alpha}(x)$.

Now the hamiltonian may be rewritten using the above phase fields. The interaction term in the Hamiltonian
can be split in two parts; one part only depends on the density of right and left movers, and gives --
 as does the kinetic energy --
a contribution quadratic in the fields $\phi_\nu$ and $\theta_\nu$, (where $\nu$ labels the eigenmodes in the diagonal basis) of the form
\begin{equation}\label{eq:Hbozon}
    H_{0}= \sum_{\nu} \int \frac{dx}{2\pi}[(u_{\nu}K_{\nu})(\pi \Pi_{\nu})^{2}+(\frac{u_{\nu}}{K_{\nu}})(\partial_{x} \phi_{\nu})^{2}]
\end{equation}
For the non-interacting system, one has $K_{\nu}=1$ for all modes, and $H_{0}$
is quadratic in the diagonal density basis which is simply
$B_{o\pi}$ (the momentum $k_{\perp}$ associated with the rungs is
either 0 or $\pi$). Another basis commonly used in the
literature is the total/transverse one $B_{+-}$. It is
related to $B_{o\pi}$ by:
\begin{equation}\label{eq:basis}
    \phi_{\mu+(-)}=\frac{\phi_{\mu o}\pm\phi_{\mu \pi}}{\sqrt{2}}
\end{equation}
where $\mu$ stands for spin or charge depending on which density
is considered.

In general $\hat{K}$ in (\ref{eq:Hbozon}) is a matrix, the form of which depends on the
basis in which the densities are expressed. For example if
we use $B_{o\pi}$  at the start of the calculation (the basis which diagonalizes the tight-binding part of the Hamiltonian),
we obtain
\begin{equation}
\hat{u}\cdot \hat{K}^{-1}= \left(%
\begin{array}{cccc}
  V_{Fo} & g_{0\pi}^{\parallel} & g_{00}^{\perp} & g_{0\pi}^{\perp} \\
  g_{0\pi}^{\parallel} & V_{F\pi} & g_{0\pi}^{\perp} & g_{\pi\pi}^{\perp} \\
  g_{00}^{\perp} & g_{0\pi}^{\perp} & V_{Fo} & g_{0\pi}^{\parallel} \\
  g_{0\pi}^{\perp} & g_{\pi\pi}^{\perp} & g_{0\pi}^{\parallel} & V_{F\pi} \\
\end{array}%
\right)
\end{equation}
and
\begin{equation}\label{eq:matrix}
\hat{u}\cdot \hat{K}= \left(%
\begin{array}{cccc}
  V_{Fo} & 0 & 0 & 0 \\
  0 & V_{F\pi} & 0 & 0 \\
  0 & 0 & V_{Fo} & 0 \\
  0 & 0 & 0 & V_{F\pi} \\
\end{array}%
\right)
\end{equation}
where $V_{Fo/\pi}$ are the Fermi velocities in the $o$ and $\pi$
bands and $g_{ij}^{\perp(\parallel)}$ are interactions between electron
densities in the $i$ and $j$ bands, with perpendicular
(parallel) spin.

In order to express the hamiltonian in a Gaussian form (Eq. \ref{eq:Hbozon}),
 which is quite convenient for the RG calculation, we diagonalize $\hat{K}$.
 This defines the  $B_{o}$ basis.
In general, $B_{o}$ is neither the
bonding/antibonding basis $B_{o\pi}$, nor the total/transverse basis $B_{+-}$.
We define the S matrix which describes the relative orientation of
the $B_{o}$ and $B_{+-}$ bases:
\begin{equation}
S=\frac{\sqrt{2}}{2}\left(%
\begin{array}{cccc}
  P_{1} & Q_{1} & 0 & 0 \\
  -Q_{1} & P_{1} & 0 & 0 \\
  0 & 0 & P_{2} & Q_{2} \\
  0 & 0 & -Q_{2} & P_{2} \\
\end{array}%
\right)
\end{equation}
One can express the parameters $P_{i}$ and $Q_{i}$ with the help of angles
$\alpha$ (for the spin part) and $\beta$ (for the charge part):
\begin{equation}\label{PQ}
\begin{split}
P_{1} &=\cos\alpha + \sin\alpha \\
Q_{1} &=\cos\alpha - \sin\alpha \\
P_{2} &=\cos\beta + \sin\beta \\
Q_{2} &=\cos\beta - \sin\beta
\end{split}
\end{equation}

\begin{figure}
\centering
\includegraphics[width=.48\textwidth, height=2.5cm]{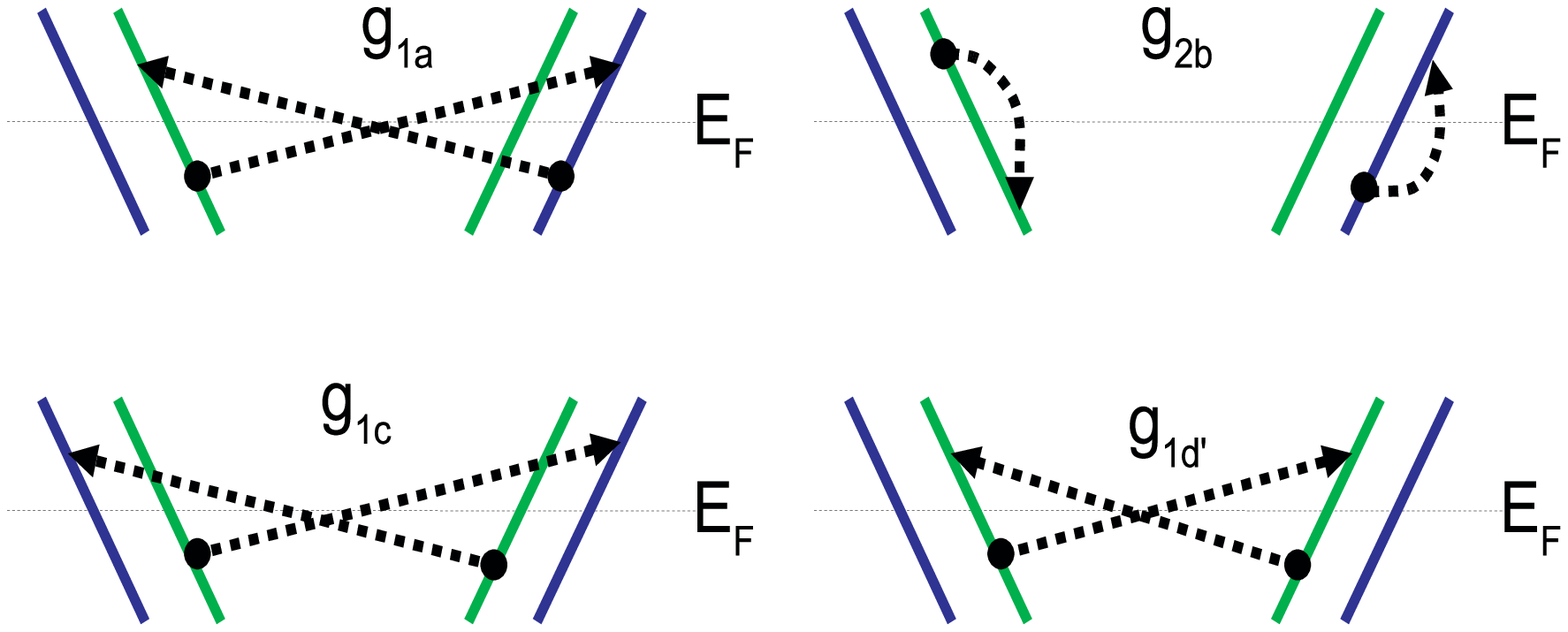}
  \caption{Diagram showing some of the scattering processes. Blue (green) lines are for carriers in the $o(\pi)$ bands.
  This illustrates the notation used for the cosine-type terms}\label{fig:proces}
\end{figure}

The remaining part of the interactions has a non-linear, cosine, form, in
bosonization language. The most convenient basis to express this contribution
is $B_{+-}$ and one finds\cite{giamarchi_book_1d, voit_bosonization_revue}
\begin{widetext}
\begin{multline} \label{eq:kosinusy}
H^{NL}_{int(1)}=-g_{1c} \int dr \cos(2\phi_{s+})\cdot
    \cos(2\theta_{c-})+ g_{1a} \int dr \cos(2(\phi_{s+})\cdot
    \cos(2(\theta_{s-})- g_{2c} \int dr \cos(2(\theta_{c-})\cdot
\cos(2(\phi_{s-})+ \\
     +g_{4a}\int
dr\cos(2\phi_{s-})\cdot\cos(2\theta_{s-})+g_{1} \int dr
\cos(2\phi_{s+})\cdot \cos(2\phi_{s-})+ g_{2}\int dr
\sin(2\phi_{s-})\sin(2\phi_{s+})+ g_{\parallel c} \int dr
\cos(2\theta_{c-})\cdot \cos(2\theta_{s-})
\end{multline}
\end{widetext}
where $\Gamma$ coefficients determine the signs of the $g_{i}$ couplings
(for instance, this gives minus signs for $g_{1c}$ and $g_{2c}$). We use the following
notation: indices 1 to 4 refer to the standard \emph{g-ology} processes for the left and right moving
carriers, letters $a$ to $d$ correspond to similar processes, when the
$o$ and $\pi$ bands labels are used instead of the left or right labels.
The relation between the $g_i$ couplings and the ones in (\ref{hubbInt}) is given in \aref{ap:couplings}.
Note that in the quadratic piece, both $g_{2}$ and $g_{4}$ -type terms need to be
included, in order to properly account for magnetic fluctuations\cite{fuseya_quasi-1Dsusc}.
Examples of interaction processes are shown in \fref{fig:proces}.
For instance, the two $g_{1d}$ terms describe events where
one right- and one left- moving fermion, both belonging to the
same ($0$ or $\pi$) band, backscatter within that band. If we
bosonize this contribution, we find two terms,
$g_{1d}\cos(\phi_{1}+\phi_{2})$ and
$g_{1d'}\cos(\phi_{1}-\phi_{2})$, instead of $g_{1(2)}$.
$g_1$ and $g_2$ correspond to the sum and to the difference of these ``1d''-type
processes respectively ($g_1=\frac{g_{1d}+g_{1d'}}{2}$,
$g_2=\frac{g_{1d'}-g_{1d}}{2}$), and $g_2\neq 0$ when the $O$
atoms are included. If the two bands were equivalent
only the $g_{1}$ process would be present.

In a standard Hubbard model, only spin perpendicular
terms are present at bare level, and the last term in Eq. (\ref{eq:kosinusy}) does not
appear at the beginning of the flow. However, Bourbonnais pointed out
\cite{nelisse_suscLL} that, during the flow towards the fixed point,
additional scatterings involving electrons with parallel spins are
generated by the RG procedure.
In our case, we are including a $V_{Cu-O}$ term so that, right from the start,
our model contains interactions between carriers with parallel spin.
$V_{Cu-O}$ gives rise to a non-linear cosine term while the other spin-parallel
processes, which are generated by the RG procedure, give contributions to the various $K$.

The $g_{4a}$ term has a non-zero conformal spin and generates two
extra couplings during the renormalization:
\begin{equation}\label{kosinusy2}
    H^{NL}_{int(2)}=
-G_{p}  \int dr \cos(4\phi_{s-})-G_{t}  \int dr \cos(4\theta_{s-})
\end{equation}
These additional terms need to be taken into account, because
they might become relevant when the other interactions
scale to zero.

\section{The Renormalization Group analysis}

\subsection{Incommensurate filling}\label{ICfilRG}

\begin{figure}
\centerline{\includegraphics[width=\columnwidth]{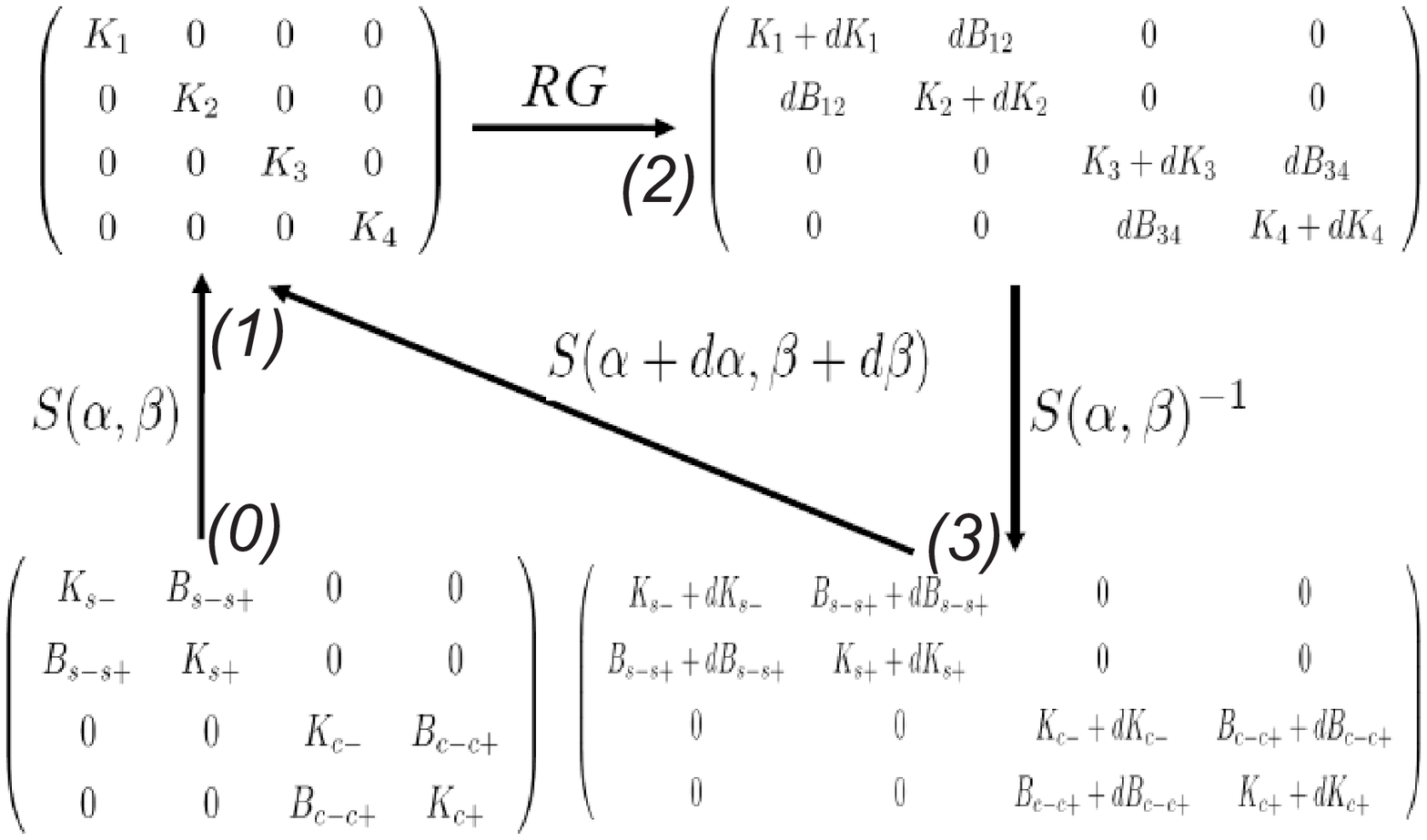}}
  \caption{Diagram showing the flow of the diagonal basis during the renormalization. The bottom row, shows the matrix in the fixed $B_{+-}$; the top row corresponds to the diagonal basis, used to write down the RG equations.}\label{fig:obrot}
\end{figure}

We start from the quadratic part of the Hamiltonian, and treat the non-quadratic part
(\ref{eq:kosinusy}) in perturbation, using a renormalization group procedure.
We compute the corrections to the correlation functions
to second order in $g$, and we incorporate them into
the LL parameters $K$. However $g$ terms are expressed in the
$B_{+-}$ basis while the quadratic part (\ref{eq:Hbozon}) is
diagonal in the $B_{o}$ basis, so the $P_{i}(\alpha,\beta)$
and $Q_{i}(\alpha,\beta)$ coefficients come into play.
As a result, 
off-diagonal terms are generated in the $K$ matrix during the RG iteration.
At this stage, $B_{o}$ is no longer the diagonal basis.
In order to fix this, the $B_{o}$ basis
has to rotate during a renormalization cycle. 
In addition to the standard RG equations for the interactions, we need to find the RG flow
of the angles $ \alpha $ (for the spin density basis rotation) and
$\beta$ (for the charge density basis rotation).
So, first we determine the corrections $dK_{1}$,...,$dK_{4}$,  $dB_{12}$, $dB_{34}$ that change the
entries of the $K$ matrix during the intial RG phase.
Next, we go back to $B_{+-}$, using the transformation $S^{-1}$; 
Since $B_{+-}$ is a fixed basis, the increments of the $K$ matrix elements give the RG step corrections
expressed in the $B_{+-}$
basis. 
This new matrix is diagonalized by the operator $S(\alpha+d\alpha,\beta+d\beta)$ 
where the angles $ d\alpha$, $d\beta$ 
depend on $dB_{\mu-\mu+}$ and $dK_{\mu-(\mu+)}$ ($\mu=c,s$). The
procedure is summarized in the diagram shown in \fref{fig:obrot}.
A detailed derivation is given in \aref{ap:RG} where, for the incommensurate case, we set all umklapp 
terms to zero in Eqs. (\ref{renormset})
and we obtain the following set of differential equations
\begin{widetext}
\begin{align}
\frac{dK_{1}}{dl} &=\frac{1}{2}[P_{1}^{2}(g_{1a}^{2}+ g_{\parallel
c}^{2}+G_{t}^{2})-K^{2}_{1}(Q_{1}^{2}g_{1a}^{2}+Q_{1}^{2}g_{1c}^{2}+P_{1}^{2}G_{p}^{2}+
P_{1}^{2}g_{2c}^{2}+\frac{1}{2}(g_{1}^{2}+g_{2}^{2})+f(P_{1})
(g_{1}g_{2}))] \\
\frac{dK_{2}}{dl} &=\frac{1}{2}[Q_{1}^{2}(g_{1a}^{2}+ g_{\parallel
c}^{2}+G_{t}^{2})-K^{2}_{2}(P_{1}^{2}g_{1a}^{2}+P_{1}^{2}g_{1c}^{2}+Q_{1}^{2}G_{p}^{2}+
Q_{1}^{2}g_{2c}^{2}+\frac{1}{2}(g_{1}^{2}+g_{2}^{2})-f(P_{1})
(g_{1}g_{2}))] \\
\frac{dK_{3}}{dl} &=\frac{1}{2}P_{2}^{2}[g_{1c}^{2}+g_{2c}^{2}+
g_{\parallel c}^{2}] \\
\frac{dK_{4}}{dl} &=\frac{1}{2}Q_{2}^{2}[g_{1c}^{2}+g_{2c}^{2}+
g_{\parallel c}^{2}] \\
\frac{dg_{1c}}{dl} &=g_{1c}\cdot[2-(P_{1}^{2}K_{2}+P_{2}^{2}K_{3}^{-1}+Q_{1}^{2}K_{1}+Q_{2}^{2}K_{4}^{-1})]-
(g_{1}g_{2c}+g_{1a}g_{\parallel c}) \\
\frac{dg_{1a}}{dl} &=g_{1a}\cdot[2-(P_{1}^{2}(K_{2}+K_{1}^{-1})+Q_{1}^{2}(K_{1}+K_{2}^{-1}))]-
g_{1c}g_{\parallel c} \\
\frac{dg_{2c}}{dl} &=g_{2c}\cdot[2-(P_{2}^{2}K_{3}^{-1}+P_{1}^{2}K_{1}+Q_{2}^{2}K_{4}^{-1}+Q_{1}^{2}K_{2})]-g_{1c}g_{1} \\
\frac{dg_{\parallel c}}{dl} &=g_{\parallel c}
(2-(P_{1}^{2}K_{1}^{-1}+Q_{1}^{2}K_{2}^{-1}+P_{2}^{2}K_{3}^{-1}+Q_{2}^{2}K_{4}^{-1}))
-g_{1a}g_{1c} \\
\frac{dg_{4a}}{dl} &=
g_{4a}(2-\frac{1}{2}(P_{1}^{2}(K_{1}+K_{1}^{-1})+Q_{1}^{2}(K_{2}+K_{2}^{-1}))) \\
\frac{dg_{1}}{dl} &=g_{1}\cdot (2-(K_{2}+K_{1}))+
P_{1}Q_{1}(K_{2}-K_{1})g_{2}-\gamma g_{1c}g_{2c} \\
\frac{dg_{2}}{dl} &=-g_{2}\cdot (2-(K_{2}+K_{1}))+
P_{1}Q_{1}(K_{2}-K_{1})g_{1} \\
\frac{dG_{p}}{dl} &=G_{p}(1-(P_{1}^{2}K_{1}+Q_{1}^{2}K_{2}))+g_{4a}^{2}(P_{1}^{2}(K_{1}-K_{1}^{-1})+Q_{1}^{2}(K_{2}-K_{2}^{-1})) \\
\frac{dG_{t}}{dl} &=G_{t}(1-(P_{1}^{2}K_{1}^{-1}+Q_{1}^{2}K_{2}^{-1}))+g_{4a}^{2}(P_{1}^{2}(-K_{1}+K_{1}^{-1})+Q_{1}^{2}(-K_{2}+K_{2}^{-1}))
\end{align}
\end{widetext}
 The equation giving the renormalization of $g_2$ measures
the influence of the $O$ orbitals.
The other two are consequences of the $g_{4a}$ term.
Note that $P$ and $Q$ depend on $\alpha$ and $\beta$ (see Eq.(\ref{PQ})),
and hence they change during the flow. 

Additional renormalization equations for the rotation of
$B_{o}$ are
\begin{align}\label{alpha}
  \frac{d\cot2\alpha}{dl}=\frac{((dK_{1}-dK_{2})\tan4\alpha+dB_{12})}{K_{1}-K_{2}}\cdot
  dl^{-1}\\
  \frac{d\cot2\beta}{dl}=\frac{((dK_{3}-dK_{4})\tan4\beta+dB_{34})}{K_{3}-K_{4}}\cdot
  dl^{-1}
\end{align}
where the equations for $dB_{12}$ and $dB_{34}$
are
\begin{widetext}
\begin{align}
\frac{dB_{12}}{dl} &= P_{1}Q_{1}((g_{1a}^{2}+g_{\parallel
c}^{2}+G_{t}^{2})-K_{1}K_{2}(g_{1a}^{2}+g_{1c}^{2}+
+g_{2c}^{2}+G_{p}^{2}))-K_{1}K_{2}h(P_{1})g_{1}g_{2} \label{B12}\\
\frac{dB_{34}}{dl} &=P_{2}Q_{2}(g_{1c}^{2}+g_{2c}^{2}+ g_{\parallel
c}^{2}) \label{B34}
\end{align}
\end{widetext}
The function $h$ is defined by
\begin{equation}
 h(P_{1})=((P_{1}Q_{1})^{2}+0.25(P_{1}^{2}-Q_{1}^{2}))^{-1}
\end{equation}
Note that we introduced $g_{1}$, $g_{2}$ -- the sum and difference
of the $g_{1d}$'s in both bands --  because the renormalization of $g_2$ involves
only $g_1$ and $g_2$.
The derivation of the renormalization equations in this
case is presented in \aref{ap:RG}. 

As was found in Ref.~\onlinecite{penc_2chain} the interband scattering
process (type $c$) renormalizes the Fermi velocities in both bands
to a common value. The additional equation taking this effect
into account is:
\begin{equation}
 \frac{d\gamma}{dl}=\gamma(1-\gamma)(g_{1c}^{2}+g_{2c}^{2}+g_{\parallel
 c}^{2})
\end{equation}
The initial value of this asymmetry parameter is $\gamma_{0}=\frac{1}{2}\tilde{\alpha}^2(\tilde{\alpha}-\frac{1}{2})^{-1}$, where
$\tilde{\alpha}=\frac{V_{Fo}+V_{F\pi}}{2V_{Fo}}$

Including this effect does not change our results, but it allows us
to determine whether
intra- or inter- band scatterings dominate for a given solution of the RG flow.

If fully spin-isotropic interactions are present in the fermionic
Hamiltonian, $SU(2)$ spin-rotational symmetry has to be
preserved during the RG flow. Some additional constrains on
the RG variables can be derived in this case. For example one of
them (for type $c$ scattering) is
\begin{equation}\label{eq:SU2}
 g_{2c}-g_{1c}-g_{\parallel c}=0
\end{equation}
Rather than using these constraints to reduce the number of RG equations, we check that they
are satisfied during the flow.

\subsection{Half filling}

If the two-leg ladder is half-filled, additional umklapp terms should be included in the Hamiltonian
\begin{widetext}
\begin{multline}\label{umklapp}
    H_{umk}=g_{3\parallel}\int dr
    \cos(2\phi_{s+})\cos(2\phi_{c+}+\delta x)+ g_{3a}\int dr
    \cos(2\theta_{s-})\cos(2\phi_{c+}+\delta x) + g_{3b}\int dr
    \cos(2\phi_{s-})\cos(2\phi_{c+}+\delta x) \\
+ g_{3c}\int dr
    \cos(2\theta_{c-})\cos(2\phi_{c+}+\delta x)
\end{multline}
\end{widetext}
Since these terms oscillate with $\delta$, their influence becomes important
only for very small doping. The extended system of differential
equations describing the RG flow has extra terms, compared with the incommensurate case,
and each of them is multiplied by a doping
dependent coefficient $J_{0}(\delta)$. The full set of equations
is given in \aref{ap:RG}.

For small $\delta$ these Bessel functions $J_{0}(\delta)$  may be approximated by one and
for large $\delta$ by zero \cite{giamarchi_umklapp_1d,tsuchiizu_confinement_spinful}. Starting from a small but non-zero doping,
assuming that  the chemical potential remains constant during the flow,
the renormalization equation that describes this Mott physics is \cite{giamarchi_mott_ref}
\begin{equation}\label{RGdoping}
    \frac{d\delta}{dl}=\delta-(g_{3\parallel}^{2}+g_{3a}^{2}+g_{3b}^{2}+g_{3c}^{2})\cdot
    J_{1}(\delta)
\end{equation}

The above equation gives an easy way to check if one is in the
insulating or in the metallic phase, and which set of RG equations
(with or without umklapp terms) is valid.  $\delta(l)$ flows to
zero for the insulator and to infinity for the metal. The value of
$\delta_{c}$ depends on the initial values of $g_{3i}$. The
description of this transition is similar to that found in
Ref.~\onlinecite{tsuchiizu_confinement_spinful}, which focused on the
confinement-deconfinement transition of two-chain systems.

For the sake of completeness, let us mention that other types of umklapp terms may appear for the two-leg ladders.
These correspond to scattering of electrons in the
bonding or antibonding bands, a process which becomes important
if one of the $k_{Fi}$ is around $\frac{\pi}{2}$. In the presence of a
large $t_{\bot}$ this condition may be fulfilled for dopings very
different from zero. For $t_{\bot}>0.1~t$ it happens somewhere
in the $C2S1$ phase. As was pointed out in the discussion of the incommensurate
case, couplings involving $\theta_{c-}$ flows then to
zero. Thus both in the charge- and in the spin- sectors, one observes the
rotation from the diagonal basis to the $k_{\perp}=0/\pi$ basis.
There are no processes competing with
this, so the only effect is the appearance of a $C1S1$ region
inside the $C2S1$ phase. These processes will not be considered in the following.

\section{Phase diagram}

Using the system of RG equations 
we determine the phase
diagram. We identify the various phases based on the behavior of
the renormalized quantities $g_{i}$. We iterate the flow up to a point when
some couplings become of order one. As usual \cite{giamarchi_book_1d}, the bosonized
form is very convenient to analyze the strong coupling case, since when coefficients in front
of cosine-like terms become large, the corresponding variables become locked.
Subsequently, one may compute the physical observable
in the ground state, by looking at the various order parameters in bosonic representation.
These operators are given in \aref{ap:operators}. Some of the operators will now have exponentially
decreasing correlations, while others will decay as power laws. The dominant phase  is the one for which
correlations decrease with the smallest exponent. It corresponds to a quasi-long range order in the ladder.

Two main factors may significantly affect the phase diagram that
was predicted for two-leg Hubbard ladders with a single orbital
per site: one is the asymmetry in the $g$ terms due to the fact
that the projections of the $Cu$ and $O$ orbitals onto the $0$ and
$\pi$ bands have  unequal amplitudes and one is the influence of
the extra parameters $U_{O}$, $V_{Cu-O}$ and $t_{pp}$.

We first investigate the impact of the asymmetry by setting
$U_{O}=V_{Cu-O}=t_{pp}=0$ and we choose a small initial values for
$U_{Cu}$ (in the range $10^{-6}\div 10^{-1}$). After this main
part we consider a few additional issues such as the
spin-rotational symmetry and the stability of the fixed points.

As in Ref.~\onlinecite{balents_2ch} we find that the parameter
which describes the behavior of the differential equations system
is $\tilde{\alpha}=\frac{V_{Fo}+V_{F\pi}}{2 V_{Fo}} $. If the ratio
$\frac{t_{\bot}}{t}$ is constant, $\tilde{\alpha}$
depends only on $\delta$: it is equal to one for half
filling (then $\delta=0$) and it reaches its maximal value when the Fermi
energy is near the bottom of the band. The parameter $
\tilde{\alpha} $ is only meaningful if the Fermi energy
crosses both bonding and antibonding bands. We restrict our
analysis to this case, otherwise one has a
single band LL. 

\subsection{Commensurate case}

Equations describing the commensurate situation are given in \aref{ap:comm}.

In this limit,  umklapp terms 
lead to  insulating phases with a gap in the charge degrees of freedom.
These states are quite similar to those presented in Ref.~\onlinecite{lee_marston_CuO}. In \sref{sec:dis-commensurable} 
We will discuss this case and also
similarities and differences with previous studies.

New and interesting physics occurs when the ladder is doped away from the commensurate case,
 and we focus on this situation
in the remaining parts of this section.
 In the incommensurate case, the asymmetry that is present when the unit cell contains two different atoms
 ($Cu$ and $O$) plays a critical role and leads to differences between the single- and multi- band models.

\subsection{The small doping case}

For small $\tilde{\alpha}$, $\cot2\alpha~\rightarrow~0$ and
$\cot2\beta~\rightarrow~0$, so the total/transverse density basis
is the eigenbasis at the fixed point. In this case, $ g_{2},g_{4a},G_{p},G_{t}$
are irrelevant.
 In the notation of Balents and Fisher\cite{balents_2ch}, this is
the \emph{C1S0} phase
 where only the $c+$
charge mode is  massless. Fields $\theta_{c-} $ and $ \phi_{s+} $
are ordered with the following
values (given mod $ 2\pi$): $ \theta_{c-}=0 $, $ \phi_{s+}=0 $.
For $s-$ the mode (spin-transverse), terms involving both
 $ \phi_{s-}$ and $ \theta_{s-}$, which are canonically conjugated become
relevant, so one observes an ordering competition between these two
fields.
The analysis of order operators presented in Appendix B shows
that SCd-type fluctuations dominate if $\phi_{s-}$ is locked at
$0$, whereas if $\theta_{s-}=0$ an orbital antiferromagnetic state (OAF) is preferred. 
In our model, SCd always dominates for repulsive
$U_{Cu}$. This prediction confirms many previous
discussions of SCd in two-leg ladders, including in the strong coupling regime
\cite{scalapino_t-Jcorrelation}, and in an inhomogeneous doping situation
\cite{wessel_inhom_doping}.

The advantage of working in bosonization language is that one can
find a reasonable quasi-classical limit of the strong coupling
fixed point. Using a semiclassical approximation for the
Sine-Gordon model \cite{rajaraman_instanton} allows one to
find the doping dependence of the gaps in the system, which up to
now was only obtained numerically . The following expression for
the soliton mass (it is the lowest lying excitation if
$0.5<K_{i}<2$) is used
\begin{equation}\label{soliton}
    m_{i}=2~\sqrt{\frac{2gu_{i}}{\pi~K_{i}}}
\end{equation}
where $u_{\nu}$ and $K_{\nu}$ are the velocity and LL parameter of
the \emph{$\nu$-th} mode (by definition we are working in the diagonal
basis), and \emph{g} is the interactions which
makes this particular mode massive. For a more detailed discussion of
gaps evaluation using RG see for example Ref.~\onlinecite{penc_numerics}.
\begin{figure}[h]
 \centerline{\includegraphics[width=\figwidth]{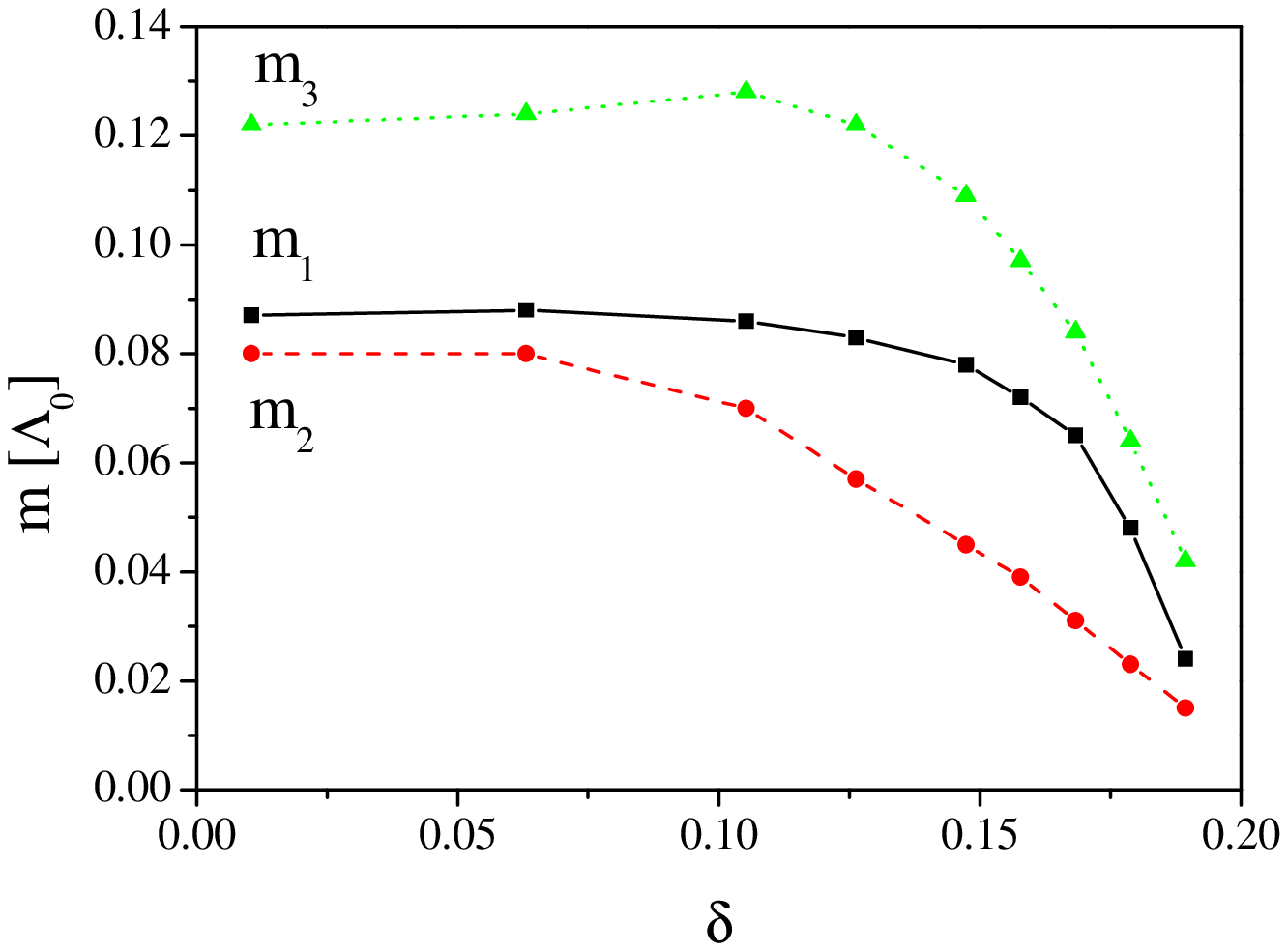}}
  \caption{Doping dependence of the gaps in the spin transverse ("1"),
   total ("2") and charge transverse ("3") density modes for the SCd phase. The $m_{\nu}$ are
   given in units of $\Lambda_{0}$, the initial energy cut-off of the RG procedure (of order $\sim 1eV$).}\label{fig:gaps}
\end{figure}
The plot shows the behavior of the masses versus doping, evaluated with the above
formula. One sees that, in the SCd phase, spin
gaps go to zero as doping increases and so does the charge
antisymmetric mode which has the largest value.
The behavior of the gaps for small doping, showing a rather slow
decay of their values is in agreement with experimental
observations \cite{kumagai_NMR_2ladder}. It is also comparable
with predictions obtained after refermionization of the problem
and mapping it onto an exactly solvable Gross-Neveu model with SO(8)
symmetry \cite{lin_so8} (but strict constraints for the
Fermi velocities -- viz $V_{Fo}=V_{F\pi}$ -- and for the ratios of the
$g$ couplings at the fixed point have to be fulfilled then).
There were also attempts to reduce the low-energy
physics of pure $Cu$ two-leg ladder to an SO(5) symmetric case
\cite{controzzi_excitation_2leg, shelton_SO(5)}. For our more
general system these conditions are usually not met,
and spurious phases may even appear if one breaks some of
the symmetries, so unfortunately we are not allowed to use these
integrable models in our calculations. However some predictions, like
the decrease of the gaps with doping and their relative magnitudes
are in complete agreement with these special cases.

It is also worthwile pointing out that the values of the two spin gaps are
always comparable, so that the approximations that are made when $m_{1}>>m_{2}$
cannot not be used here to calculate the physical properties of our model.
This behavior pertains to the range $\delta<\delta_{c1}$; upon approaching  $\delta_{c1}= 0.2$
from below, gaps tend to close. As we will show next, a different phase,
\emph{C2S1}, emerges for $\delta>\delta_{c2}$. The intermediate range $\delta_{c1}<\delta<\delta_{c2}$ will be discussed separately, when we examine
the transition from the $C1S0$ to the $C2S1$ phase.

\subsection{The large doping case}

\begin{figure}[h]
 \centerline{\includegraphics[width=\figwidth]{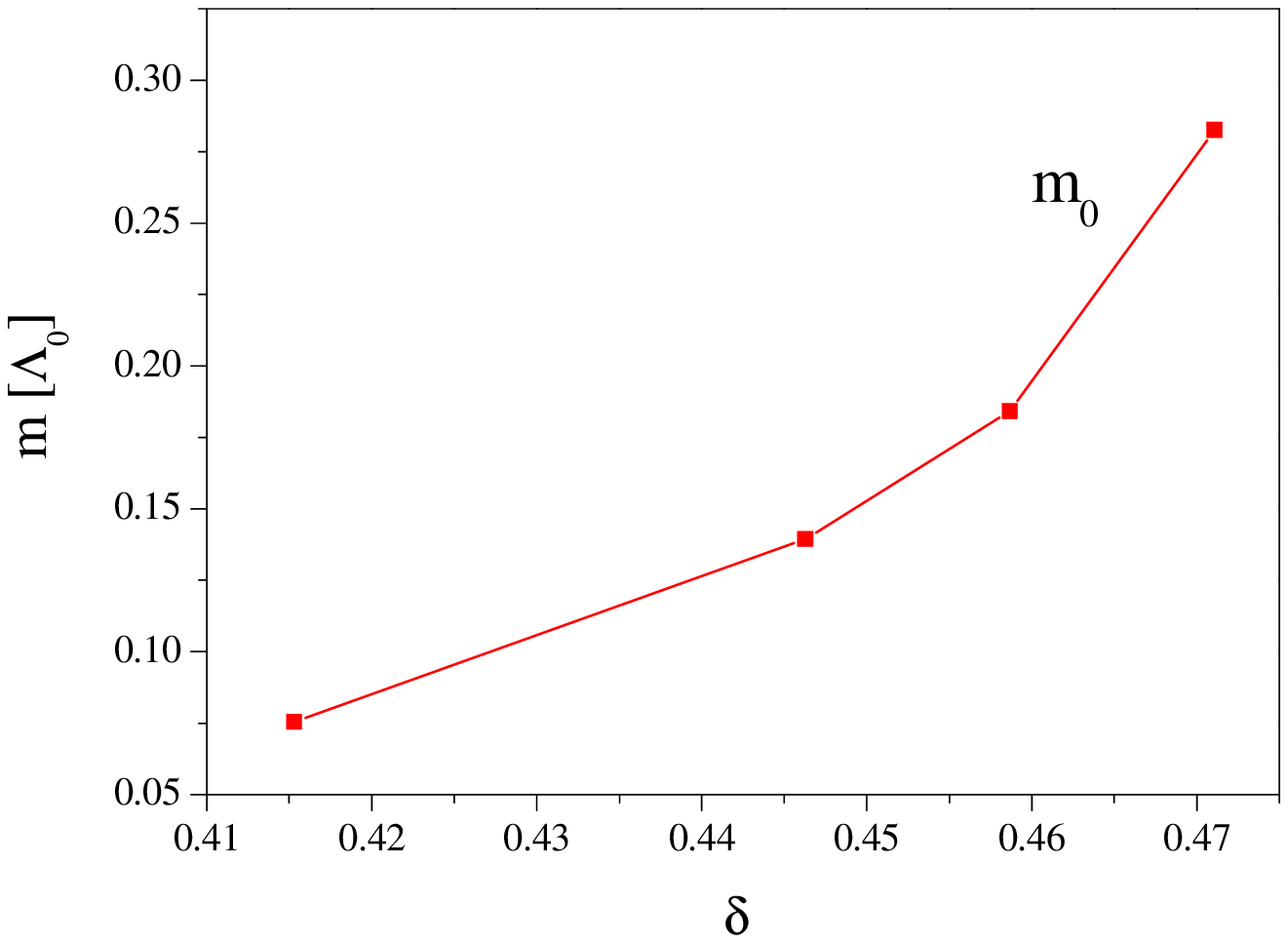}}
  \caption{Doping dependence of the gap for the \emph{`'o"} band
  spin mode. Note that the gap is given in units of the cut-off $\Lambda_{o}$, the value of which decreases quickly when one approaches the bottom of the upper band}\label{fig:gaps2}
\end{figure}

If the asymmetry between the bonding and antibonding bands is larger,
 $\cot2\alpha~\rightarrow~\infty$ and
$\cot2\beta~\rightarrow~-\infty$, signalling that  $B_{o\pi}$ is the eigenbasis
 for both the spin and charge modes. The RG flow converges very quickly
to that fixed point for large dopings, typically when $\delta>0.41$.
This corresponds to
$\tilde{\alpha}= 4.2$, a value that agrees with that
found previously in Ref.~\onlinecite{balents_2ch}.
Interactions are not able to renormalize the ratio of the
band Fermi velocities $\gamma$ to one anymore, which confirms that the $B_{o\pi}$
basis is relevant for this regime.
In the large doping phase, only $ g_{1} \simeq -g_{2}$ are
relevant.
If one takes into account the rotation of the diagonal basis, which occurs as
 $\cot2\alpha$ varies, it appears that this flow produces
only one massive spin mode, and we get the $C2S1$ phase predicted by Balents and Fisher
\cite{balents_2ch}; the
interaction term which causes this behavior in our case is identical to
theirs, once expressed in current density
formalism. Using the same method as the one described for low
dopings we are able to evaluate the doping dependence of the gap
of strongly doped ladders. The result is shown in
\fref{fig:gaps2}.

One can easily identify the nature of the $C2S1$ phase, in
bosonic field language: because the diverging interaction is
$g_{o}=g_{1}-g_{2}<0$, the slowest decay of correlations is
observed for the CDW operator in the bonding (\emph{"o"}) band.

For large enough dopings ($\delta>0.41$), a gap will open, even if one starts from very small
bare values of the interactions. In
the range $0.28<\delta<0.41$, angles still flow to the fixed point limits
$\cot2\alpha(\beta)~\rightarrow~\infty$, but $g_{1}$ and $ g_{2}$
grow very slowly. One needs to choose larger initial values
of the bare interactions (but still
smaller than the hopping \emph{t}) and/or assume
that one is close to the $C2S1$ region (starting from large
$\cot2\alpha(\beta)$) to find the gap exactly in the relevant spin
mode. We conclude that the $C2S1$ phase exists in the entire range
$\delta>\delta_{c2}=0.28$ and that, when $0.28<\delta<0.41$, $g_{1}$ and $ g_{2}$ are very
weakly relevant and thus very sensitive to higher order corrections.

Previously, the existence of a $C2S2$ massless phase was predicted
very close to the bottom of the band, (i.e when $\tilde{\alpha}$ becomes quite large).
Our calculation, however, shows that the $C2S1$ phase remains stable in that limit.
The reason for this difference stems from the choice of initial conditions in
Ref.~\onlinecite{balents_2ch}.
For single orbital ladders, when \emph{only} on-site Hubbard interactions are included,
the initial $g_{2}$ is accidently zero. In our case, the presence of $O$ orbitals always implies
a non-zero initial $g_{2}$.
This says that the $g_{2}$ term drives the transition, for very large
$\tilde{\alpha}$.

At the bottom of the band the dispersion is quadratic, so the
bosonization procedure, which requires a linear spectrum around
the Fermi points, is not valid. The calculation is done using conventional
diagramatic techniques, and it confirms the stability of the $C2S1$ phase with
the same relevant coupling as found before.

\subsection{Quantum critical regime between the $C1S0$ and $C2S1$ phases}

We now turn to the intermediate regime $\delta_{c1}<\delta<\delta_{c2}$.
Our key finding is that in this range, a new, massless, phase exists that
had not reported previously, because it is found
in the asymmetric limit, i.e when the unit cell contains both $Cu$ and $O$ orbitals.

When one approaches the range $\delta\in(0.2;0.28)$ either from below or from above, gaps appear 
to go to zero (see Figs. \ref{fig:gaps} and \ref{fig:gaps2}).
For $\delta_{c1}<\delta<\delta_{c2}$, one has a line of critical points where the phase
is totally massless ($C2S2$).
Strong fluctuations, in particular near the critical end points $\delta_{c1}$ and $\delta_{c2}$, 
cause poor convergence of the RG differential equation system. What is more, the angles Eq. (\ref{alpha}) 
vary significantly in a narrow range of $l$.
Using controlled approximations, we obtain an analytical solution that reveals the behavior 
of the system in this phase: we approach the critical end points from the massive phases;
we only keep the dominant couplings, which give us a simplified
system of equations. Next we analyze the equations describing the
angle rotations, and look for the range were derivatives become large, which takes place close to the fixed points.
 This gives us a
condition for the divergence of $\cot2\alpha(\beta)$. Once the fixed
point is known, one may simplify further the RG
differential system. Now computing the RG exponent of each
coupling is straightforward and enables us to find those couplings which
remain relevant within the range of interest.

Let us first consider $\delta_{c2}=0.28$. This point
corresponds to the initial value $\cot2\alpha=1$.
A numerical solution shows that the signs of
$(dK_{1}-dK_{2})$ and $B_{12}$ are the same and positive whereas
the sign of $(K_{1}-K_{2})$ is negative. From this simple analysis
we infer that below this value $|\cot2\alpha|$ decreases to zero
and that above, it increases to infinity. Now,  $g_{1}$ and $ g_{2}$
are only relevant when  $(K_{2}+K_{1}))-P_{1}Q_{1}(K_{2}-K_{1})<2$.
$K_{2}$ needs to decrease strongly for this condition to be fulfilled and
it is necessary to have nonzero values of $f(P_{1})$ and $Q_{1}$
at the fixed point. This condition corresponds to
$|\cot2\alpha|\rightarrow\infty$, so one sees that below
$\delta_{c2}=0.28$ the $g_{1}$ and $ g_{2}$ couplings cannot be
relevant.

The analysis pertaining to $\delta_{c1}=0.2$ is less straightforward.
It involves $|\cot2\beta|$ and, because $(dK_{3}-dK_{4})$
and $B_{34}$ have opposite signs, it is harder to get the flow correctly.
The transition between going to zero and diverging takes place when
the absolute value of the two terms are equal. A detailed analysis of the angle dependent part of
$d\cot2\beta$ shows that this happens for $\delta_{c1}=0.2$. When
$|\cot2\beta|\rightarrow\infty$, then $K_{4}^{-1}$, which is much
larger then one, influences the renormalization of the
$\cos\theta_{c-}$ coupling on equal footing with $K_{3}^{-1}$. This
is the reason why these interactions are not relevant anymore.

The above first order RG analysis was done
in the vicinity of the critical end points and proves that when
$\delta\in(0.2;0.28)$ all interaction terms which are relevant
outside are irrelevant inside this range. We have confirmed the
above simplified analysis by performing a numerical analysis of the full set of equations
 which shows that no other coupling is
relevant. We see that a $C2S2$ phase is present
between the $C1S0$ and $C2S1$ phases. Hints for the possible existence of such state came from numerical studies \cite{nishimoto_DMRG_C2S2} or from some special models of ladders
\cite{baruch_bondallign_C2S2, fukui_massless} with  specific types of geometries, but we give here a direct proof of the existence of this phase for a generic ladder.

The $C2S2$ phase is a LL where $B_{o\pi}$  is
the fixed point eigenbasis for the charge modes and $B_{+\-}$  that for the spin modes.
As far as the charge modes are concerned, $K_{4\equiv o}$ is significantly smaller than one,
while $K_{3\equiv \pi}$ is very close to one at the fixed point.
The spin parameters are both close to one because of the spin rotational symmetry.
Thus one expects
that correlation functions of band density fluctuations of the form $c_{+/-,o}^{\dag}\sigma_{i}c_{+/-,o}$ have
the slowest decay. Logarithmic corrections need to be
evaluated, owing to the presence of a (single) marginal coupling  $g_{o}=g_{1}-g_{2}>0$. 
They show that a SDW  within the \emph{o-band} (\emph{SDW(o)}) is dominant.


\subsection{The influence of $ U_{O} $ and $V_{Cu-O}$}

Sofar, we have only discussed changes that stem from the presence of  $O$
orbitals in the structure. We now turn on the interactions involving the $O$ atoms
 -- $ U_{O} $ and/or $V_{Cu-O}$ -- and probe whether these additional terms
affect or not the phase diagram that we have found previously. In the
following, we assume that these interactions do not
generate new types of terms in bosonization language, but that they
modify the initial parameters of the flow (for a detailed
discussion of \emph{V}-type terms, see for example
Ref.~\onlinecite{voit_extendHubb}).

In the $C1S0$ phase, SCd becomes less stable if large
$U_{O}$ or $V_{Cu-O}$ are present, but it has always
a lower free energy than the OAF phase.
 When both $ U_{O} $ and $V_{Cu-O}$ are present, they seem to have competing effects.
 One would need to assume a very large attractive bare $V_{Cu-O}$
($V_{Cu-O}<-3.6 U_{Cu}$) in order to stabilize a phase different from SCd.
 It would be a SCs phase with $\phi_{s-}=0$, $\phi_{s+}=0$, $\theta_{c-}=\pi/2$ and  
it would be very robust, even if the Fermi level
approaches the bottom of the $\pi$ band. The existence of this
phase, generated by  $V_{Cu-O}$, was first pointed out in Ref.~\onlinecite{lee_marston_CuO}. The
discussion of Ref.~\onlinecite{lee_marston_CuO} pertains to the half
filled case. The nature of the phase transition between the two
superconducting phases in ladders was described in detail in
Ref.~\onlinecite{tsuchiizu_2leg_firstorder}, so we are not going to discus
this point. In the physical range of values of bare $V_{Cu-O}$, one does not expect SCs to dominate.

In the $C2S1$ phase, $U_{O}$ and $V_{Cu-O}$
do not change the results significantly. Their main influence is
that they make the gap smaller. It is to be expected, since
the CDW in the "o" band  has an overlap with $O$ atoms sitting between two $Cu$
and introducing electron repulsion on $O$ atoms
makes the CDW less stable. For very large attractive $V_{Cu-O}$
the SCs phase re-enters.

In the $C2S2$ phase, increasing $V_{Cu-O}$ has little effect on
$\delta_{c2}$ but shifts $\delta_{c1}$ to higher values. For
${{V_{Cu-O}}\over {U_{Cu}}}>1$ the quantum critical line still
exists and an unphysically large ratio of $\approx 5$ would be
required to suppress the massless phase and to observe a reentrant
$C1S0$ phase with superconducting fluctuations. The $\delta_{c1}$ phase boundary is
not affected by $U_{O}$ or by $V_{Cu-O}$.

\section{Discussion and consequences}\label{discussion}

In this section, we discuss our findings in connection with previous work done on ladders.
 We also show that the $C2S2$ and $C2S1$ phases possess
an orbital current quasi long-range order and we compare our result
 with other proposals of current patterns for cuprates.

\subsection{Differences with the single band case}

In the derivation of the RG equations using current algebra, total
particle-hole symmetry was assumed. Yet, it was shown \cite{wu_2leg_firstorder} that V-type
interactions, for instance, can generate terms which break this symmetry at
the beginning of the flow. They generate the following terms
\begin{itemize}
    \item a sine interaction term $g_{2} \sim g_{1o}^{\perp}-g_{1\pi}^{\perp}$
    \item  interactions such as $g_{2o}^{\perp}-g_{2\pi}^{\perp}$
    and $g_{1o}^{\parallel}-g_{1\pi}^{\parallel}$, of the form $\nabla\phi_{s+}\nabla\phi_{s-}$, which are
    included in the definition of the non-diagonal part of the $\hat{K}$
    matrix (see Eq. (\ref{eq:Hbozon})); this implies, for instance, that $P_{1}\neq P_{2}$
    \item $g_{4}$-type interactions generate different
    velocities for the spin- and charge modes (\emph{per se} this is not a relevant
    perturbation but it enhances the impact of the other two contributions).
\end{itemize}
When $O$ atoms are included between $Cu$ atoms, even
if \emph{only} $U_{Cu}$ is present, the bare
$g_{0}\sim\frac{\lambda_{ao}^{4}}{V_{Fo}}U_{Cu}$ is different from
$g_{\pi}\sim\frac{\lambda_{a\pi}^{4}}{V_{F\pi}}U_{Cu}$ and
particle-hole symmetry does not hold anymore (in the limit
$E\rightarrow\infty$, one has
$\lambda_{ao}^{4}-\lambda_{a\pi}^{4}\sim E^{-1}$ and $V_{F}\sim
E^{-1}$ so these two effects cancels out and $g_{o}-g_{\pi}$ is still $\sim O(1)$).
This shows that it is quite important to include the oxygen atoms in the description of the two-leg
ladder.

The system of RG equations that we have derived does not impose such
particle-hole symmetry constraint, and hence it may flow to a new fixed point which corresponds to
the $C2S2$ phase. At the fixed point, $B_{+-}$ is the
diagonal basis for the spin modes and $B_{o\pi}$ is the diagonal
basis for the charge modes. It should be emphasized that for all
other phases (which had been found previously for single orbital ladders),
the diagonal basis at the fixed point is the same for the spin and for the charge modes.
The presence of the three bands thus allows the symmetry between spin and charge bases to be relaxed during
the flow, and is instrumental in stabilizing the $C2S2$ phase. For the case of a single band, 
Ref.~\onlinecite{emery_stability}
pointed out that two additional considerations could lead to a significantly modification
 of the phase diagram obtained
in Ref.~\onlinecite{balents_2ch}, using a weak coupling perturbative approach.
 The first one was the inclusion of
all interactions, not simply the relevant ones, the second one was the stability of the fixed points.
 For example in
Ref.~\onlinecite{emery_stability}, it was argued that the stability
of the $C2S1$ phase was compromised, because of ``a spin proximity
effect''. However this $C2S1$ phase was found in DMRG numerical studies
\cite{park_2ch_dmrg}. In our calculation it is important to note that all
possible interactions were taken into account, and we did not impose any \emph{a priori} symmetry.
The presence of the $C2S1$ phase, that we do find in our calculation, is thus intimately connected with the
rotation of the spin basis towards the fixed point $B_{o\pi}$ eigenbasis.

\begin{figure}[h]
  (a)
  \centerline{\includegraphics[width=0.62\columnwidth, angle=-90]{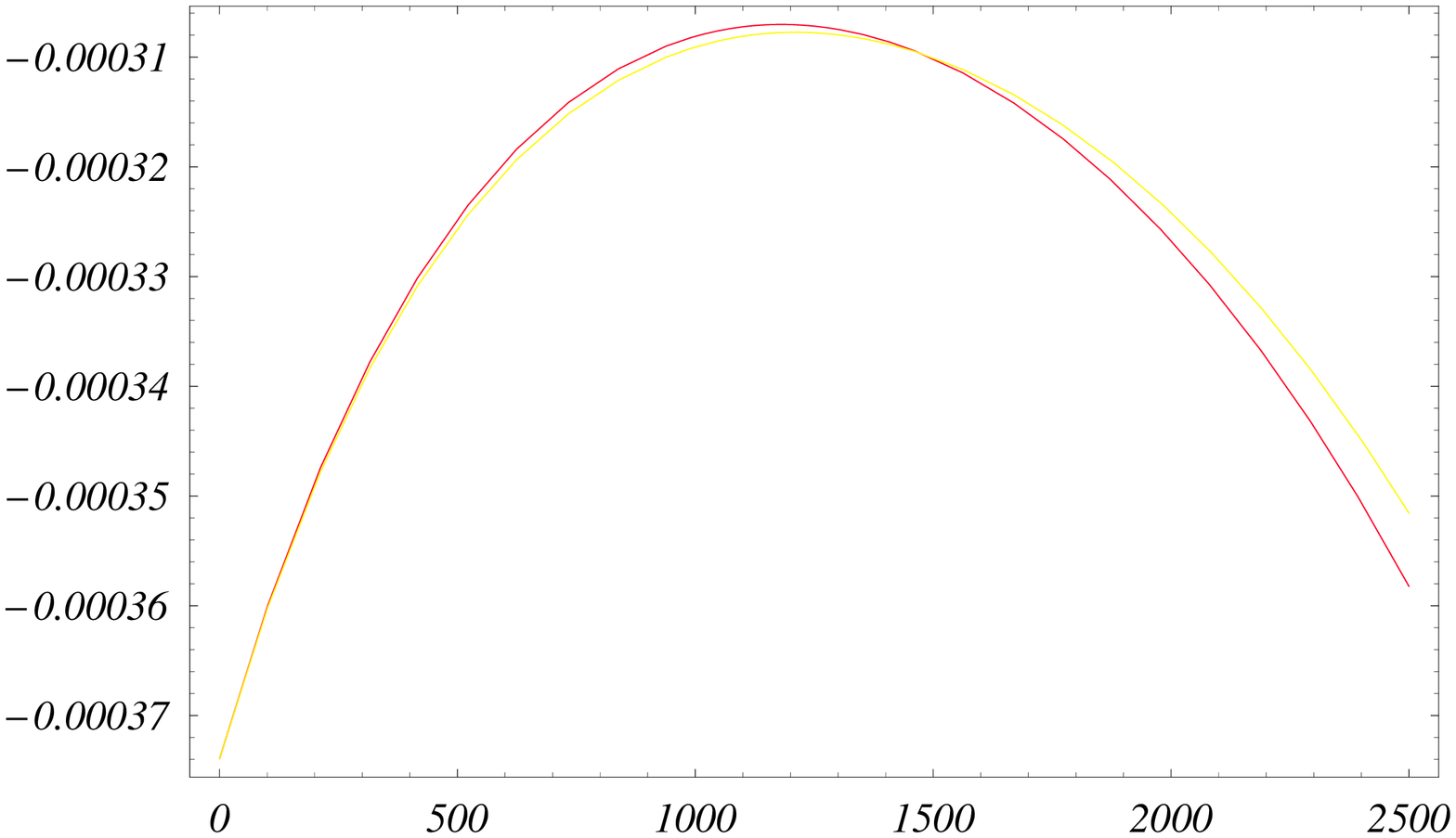}}
  (b)
  \centerline{\includegraphics[width=0.62\columnwidth, angle=-90]{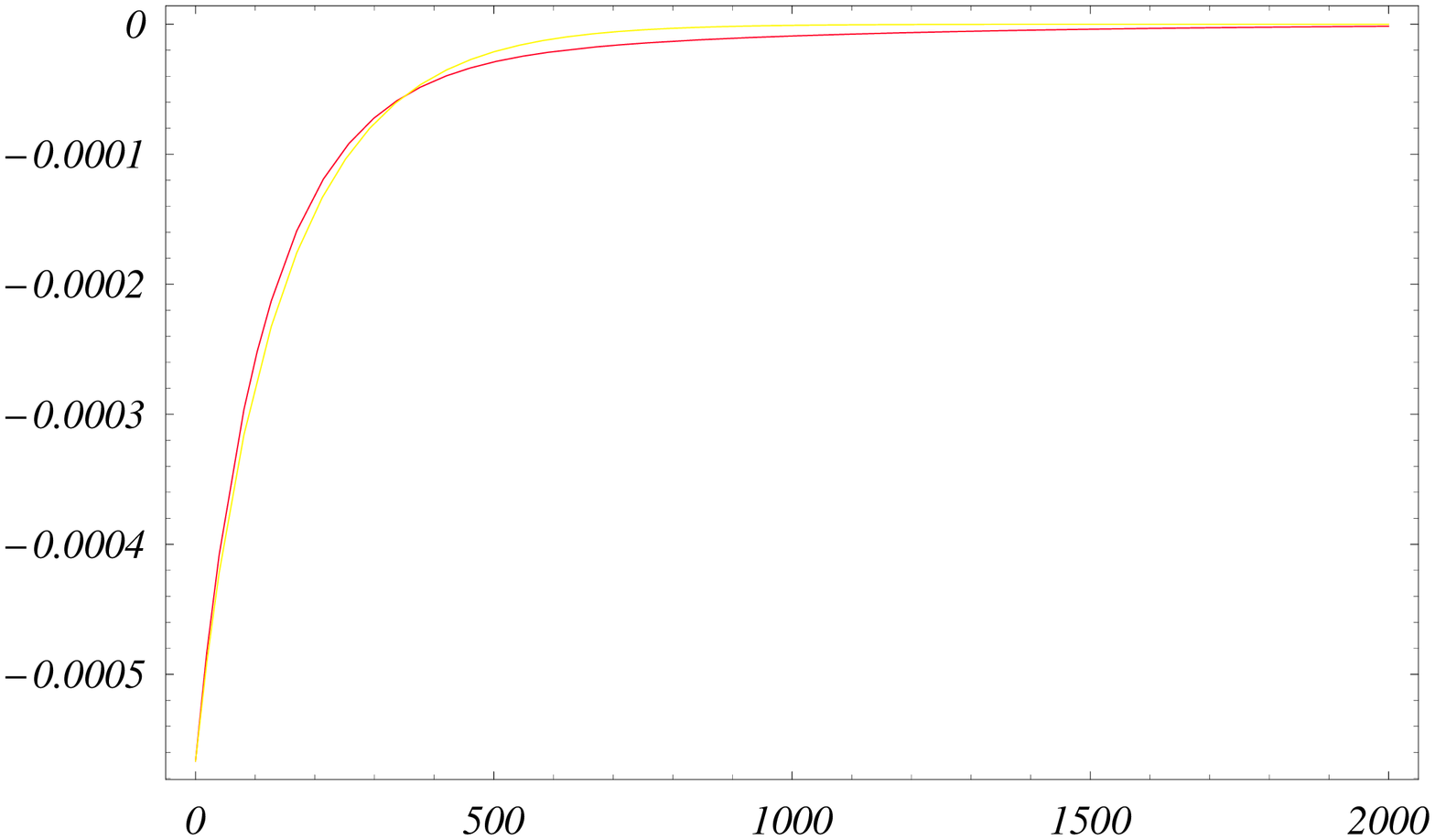}}
  \caption{Evolution of $g_{2c}$ (yellow) and $g_{1c}+g_{\parallel c}$ (red) versus number of RG steps,
during the flow. It shows that spin rotational symmetry is preserved
  (a) small $\alpha$ phase (b) large $\alpha$ phase}\label{fig:spin_rot}
\end{figure}

At each step, we monitored the spin rotational invariance of the Hamiltonian
through Eq. (\ref{eq:SU2}) to check
that our equations were producing a reliable flow.
The result,  for the case of small
as well as large dopings, is displayed in Fig.~\ref{fig:spin_rot}.
In addition, in the $C2S1$ phase, there is one single massless spin mode, so that its  $K$
parameter must remain equal to one during the flow; we verified that
this property does hold.




\subsection{Half filling and close to half filling}\label{sec:dis-commensurable}

At half filling, the charge symmetric mode
becomes massive. All modes are gapped and spatial correlations decay exponentially.
This is due to three relevant umklapp
couplings. The spin and charge transverse modes are locked into the
same minima as before, and the transition only affects the total charge
mode. One may view this transition as a quantum
order-disorder Ising type. At half filling the dominant phase is
the quantum disordered D-Mott phase, which, upon doping, turns into SCd, its
dual counterpart. For large attractive \emph{V}, an S-Mott phase, the dual
counterpart of SCs, dominates at half filling. When we vary the strength of the interactions,
the boundaries between these two phases look similar to those found for incommensurate fillings.

The half-filled case for the $Cu$-$O$ ladder was discussed
in Ref.~\onlinecite{lee_marston_CuO}, both in the weak and in the strong
coupling limits. For weak interactions, we may
directly compare their results with ours. They used  current
algebra to treat the low energy physics of
ladders with and without outer
oxygens (five and seven atoms in the unit cell respectively).
In the latter case, a spin-Peierls phase (BDW in Appendix.B)
dominates, whereas a D-Mott phase is favored, in the former case. The authors
claim that this difference is due to a larger leg to rung anisotropy when outer oxygens are not present.
 The outer oxygens
were taken into account in our model but we nevertheless find a D-Mott
 phase. More generally, our entire phase
diagram is very similar to their ``five
orbital'' case. 
A possible reason for this discrepancy
could be that

their $t_{pp}$
is barely less than the $Cu$-$O$ hopping amplitudes.
In our calculations $t_{pp}$ is much smaller, in accordance with LDA studies \cite{Muller-Rice_LDA} 
and with experiments.
Recall that, as was described in Section II, whenever $t_{pp}>0.5t$ (or $t_{\bot})$
non bonding p-orbitals become relevant degrees of freedom, and these were not included in our model.
Similarly a large initial value of the nearest-neighbor interaction $V$ 
causes an exchange of the weight of the \emph{d} and \emph{p} orbitals in
the lowest lying bands during the RG flow. This limit is beyond the range of validity of a
simple RG approach.

The strong coupling case
(the so called charge-transfer regime) is important,
because, for real inorganic materials,  $U$ is usually of order 5t.
Still,
two features of the weak coupling regime remain valid
in strong coupling: one is that spin-charge separation holds and two is
that in the SCd phase (for instance) there is still an exponential
decay of DW operators. A connection between the phase diagrams
of these two regimes is often suggested in the literature.

For instance in the case of $Cu$-$O$ ladders close to half filling
(the strong coupling case discussion in Ref.~\onlinecite{lee_marston_CuO}) a
\emph{t-J} approximation was used. It gave a uniform
phase -- related to D-Mott --  in a broad region of positive
$U_{Cu}$-$V_{Cu-O}$ phase space. For large attractive $V_{Cu-O}$,
a phase with holes localized around copper atoms is found,
probably connected to our SCs ordering. Our phase diagram matches the above description.
The SCd phase, which we find close to half filling, is clearly seen in numerical studies.
The $C1S0$ phase was connected with this
type of ordering in Ref.~\onlinecite{scalapino_t-Jcorrelation} were it was also shown
that the gap preventing a DW-type ordering decreases upon increasing the doping. In Ref.~\onlinecite{gros_DMRG_phases}
it was found that the region where this phase is stable can be extended
up to U=4t. In the t-J model, the rigidity of SCd with respect to a
finite difference in the chemical potential of the two-legs was also established \cite{wessel_inhom_doping}. 
The same type of ordering (rung singlet) also dominates at half filling, for a special choice of
parameters giving an SO(5) symmetry \cite{marston_oaf_ladder}, since in that case one may solve the model exactly.
All these results
were obtained for single band ladders; quantitative differences occur when oxygen atoms are included
 in the unit cell, and these were analysed in a numerical study \cite{jeckelmann_DMRG}.

Few studies were devoted to the intermediate and large doping regimes; we discussed the $C2S1$ phase (see above),
and, as far as the
$C2S2$ phase is concerned, a DMRG study \cite{nishimoto_DMRG_C2S2} 
suggested the existence of such gapless phase well inside the bands for a zig-zag ladder;
the occurrence of a massless phase in the strong interaction limit would be certainly remarkable.

\subsection{Orbital current patterns at intermediate and large dopings}

In the previous sections, we had set $t_{pp}=0$. We now  assess
the influence of this hopping term on the phase diagram. As long as $t_{pp}<\emph{0.5t}$, our RG
method remains valid, and only the initial parameters are
changing with $t_{pp}$. As we increase $t_{pp}$, we note that  $\tilde{\alpha}$ increases for a given doping, but that both $\delta_{c1}$ and $\delta_{c2}$
are decreasing. For $t_{pp}=0.5$, 
their values are about half that quoted for $t_{pp}=0$. A
negative $t_{pp}$ (the electron-doped case) has the opposite
effect. This influence of $t_{pp}>0$ can be understood as an
increased asymmetry between bands.

\begin{figure}[h]
  \centerline{\includegraphics[width=\figwidth]{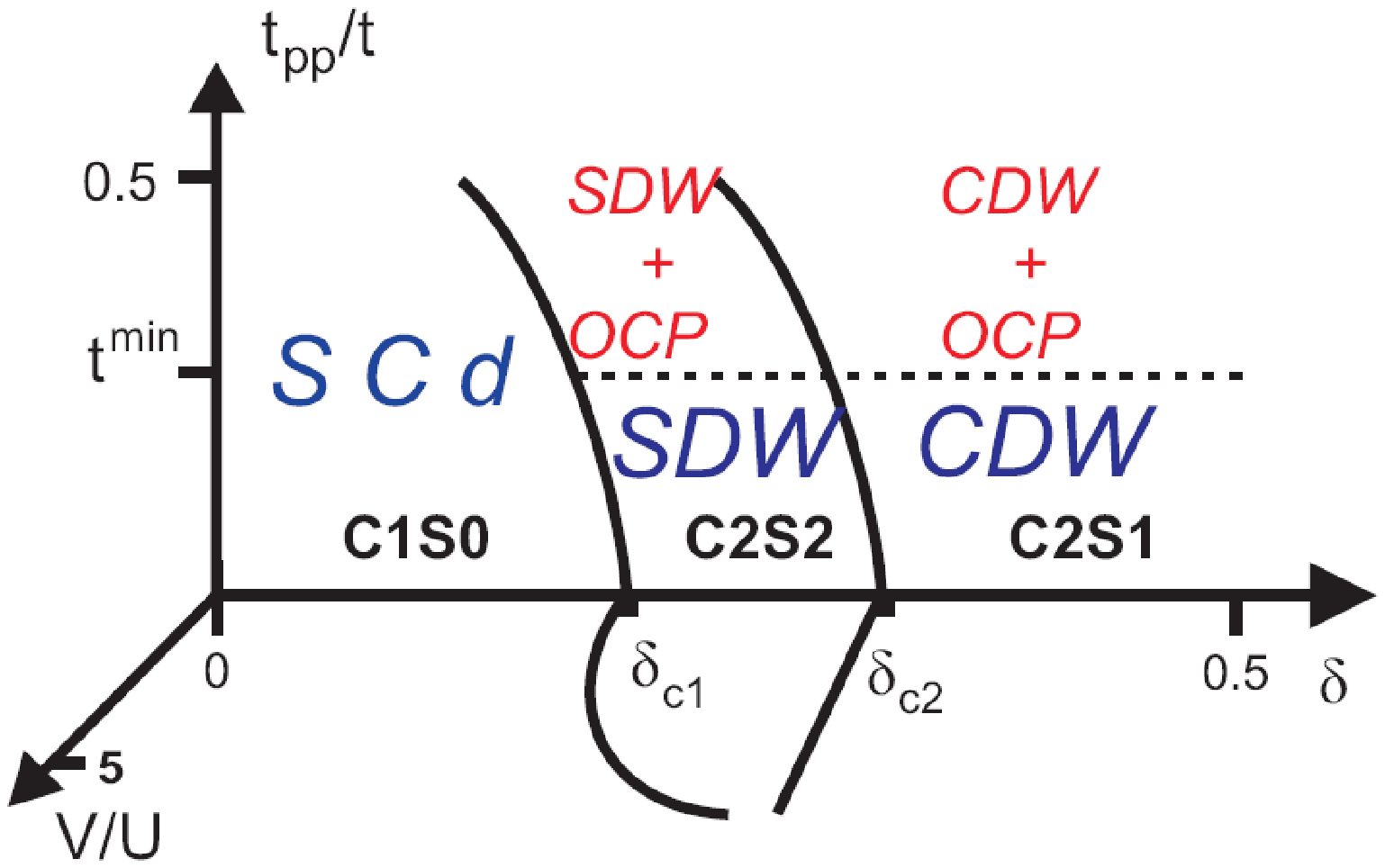}}
  \caption{Phase diagram of two-leg Hubbard ladders versus doping
  for $U_{Cu}>0$. $\delta=0$ corresponds to
   the half filled case; umklapp terms which open up a gap in
  the charge symmetric mode are not included here;  \emph{DW+OCP} denotes
   phases with orbital current pattern (OCP) on top of a spin- (SDW) or a charge- (CDW) density wave.}\label{fig:dopin}
\end{figure}

The phase diagram,  which summarizes our study of the $Cu$-$O$ ladder for carrier concentrations
between half filling down to the bottom of the ``upper'' band, is shown in Fig. \ref{fig:dopin}.

A spectacular effect of $t_{pp}$ is that it leads to new types of current loops, involving oxygen sites; 
the range of parameters where  orbital current patterns (OCP) dominate is seen   
in Fig. \ref{fig:dopin}. 


A finite $t_{pp}$ allows direct current flows between oxygen
atoms, giving rise to additional patterns, enclosed inside the
elementary cell. One of these preserves the mirror symmetry on the $\sigma$ axis (the axis parallel to the chain direction
and passing through the mid-rung oxygens), and it
is of special importance. This is because we have shown that, at least for moderate dopings, operators in the $o$
band dominate.
The current operator between two atoms "a" and "b" is defined as
$j_{a-b}=\sum_{\alpha,\alpha'}\psi_{\alpha a}^{\dag}\psi_{\alpha'
b}-\psi_{\alpha' b}^{\dag}\psi_{\alpha a}$ (we sum over band indices) and the total current
pattern operator is given as a sum of currents on each bond. For symmetry reasons, if the current pattern
has a mirror symmetry along $\sigma$,  then the total current
operator has the form of a particle-hole fluctuation in the $o$
band.

\begin{figure}
 \centerline{\includegraphics[width=\figwidth]{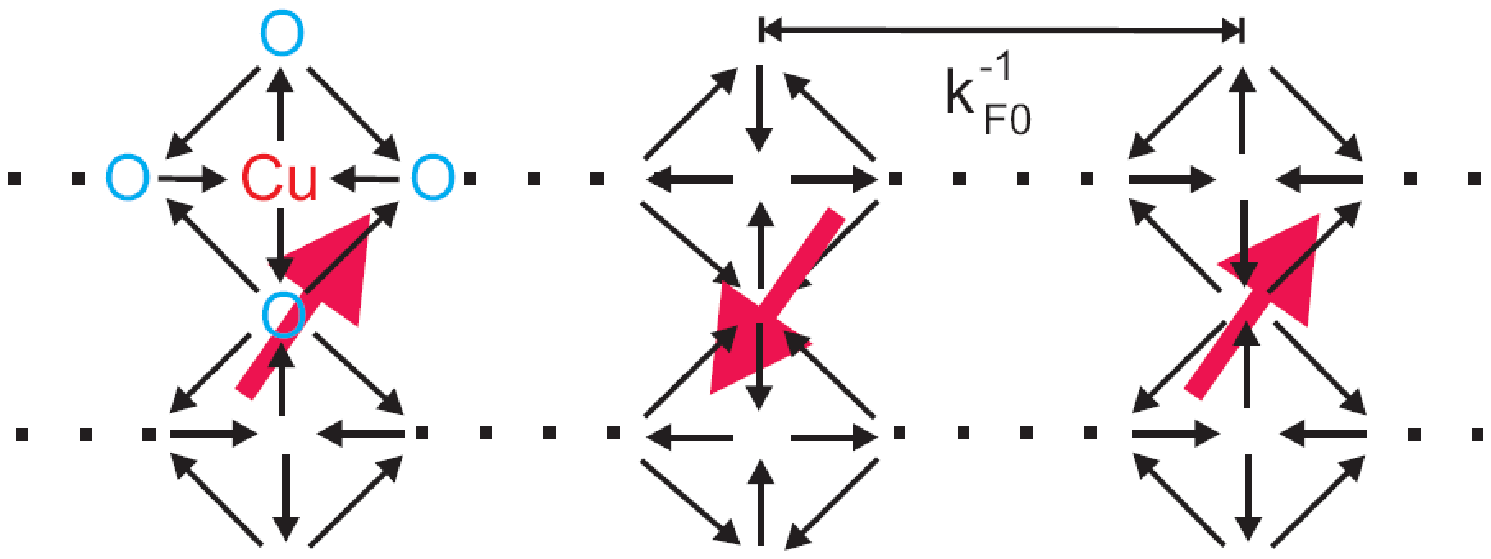}}
  \caption{Current pattern in the $C2S2$ phase. It has a mirror symmetry with respect to the $\sigma$
 axis plus an additional SDW. The modulation has an incommensurate spatial periodicity ${2k}^{-1}_{F0}$}
\label{fig:Varma}
\end{figure}

Two conditions must be met in order to get a
dominant contribution: the pattern must form closed loops originating from and ending at
 $Cu$ atoms and it has to possess a  mirror symmetry with respect to
the plane containing $\sigma$ and perpendicular to the plane of the ladder. One of the current patterns
preserves both of the required
constrains and, since it is similar to a configuration proposed by Varma, 
we call it \emph{"VarmaI"-type} 
(Fig. \ref{fig:Varma}). 
Then, because the current operator has the same dependence on phase fields as the DW operator,
these fluctuations have the same power law decay.
Computing their amplitude will tell us which type of order dominates.


In the large doping regime ($C2S1$), we compare the amplitude of the
\emph{"normal"} CDW and of the OCP+CDW. The
amplitude of the latter is found by summing
current operator contributions for loops with one $Cu$ and two $O$ atoms. We
use the mirror symmetry and add first equivalent pairs of
 currents. Each of these pairs gives a contribution
proportional to $t_{ij}Im(\lambda_{a\alpha}\lambda_{bi\alpha})$ or
$t_{ij}Im(\lambda_{bj\alpha}\lambda_{bi\alpha})$, where $t_{ij}$
is the hopping parameter between the relevant atoms. We
emphasize once again that, in the  single ($Cu$) orbital case, $\lambda_{a\alpha}=1$
so that the current operator between $Cu$ atoms has the usual
interband form. It is the presence of oxygens that gives
$Im(\lambda_{a\alpha}\lambda_{bi\alpha}\neq 0$, allowing
the geometry of a \emph{Varma-type} pattern to appear in the
theory.

\begin{figure}
 \centerline{\includegraphics[width=\figwidth]{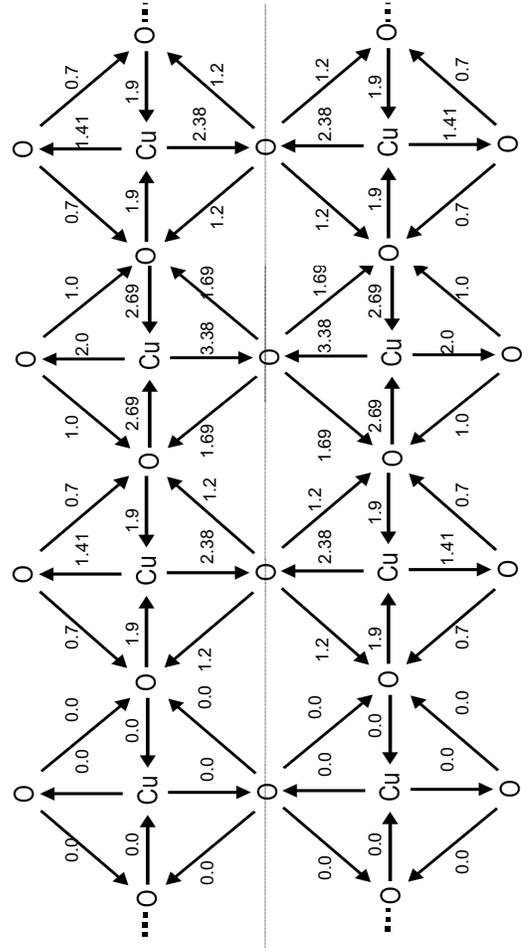}}
  \caption{Current amplitudes within the o-band,
given in $t_{pp}$ units. For this particular doping, the pattern has an 8-cell periodicity;
 The figure shows half a period (the other half is simply obtained by repeating the amplitude pattern 
and reversing all the signs) }\label{fig:pattern}
\end{figure}

A numerical calculation shows that these quantities are
of order one and change only by a few percent when the doping
increases from 0.25 to the value of $\delta$ at the bottom of the band. The result of this
procedure (the amplitudes of the currents determined by the products of $\lambda_{ij}$ coefficients)
  is shown in Fig. \ref{fig:pattern}; since it is easier to visualize a commensurate pattern, we chose 
$\delta=0.9$ such that $k_{Fo}=1/4$ (note that only the ``0'' band would cross the Fermi level for such
a large value of the doping). 

Due to current conservation, the weakest link between
atoms determines the maximum value of the current. It is clear that
the magnitude of $t_{pp}$  determines whether or not
the OCP+CDW state may exist. Since the
total amplitude is proportional to
$2t_{pp}\sum_{ij}Im(\lambda_{bj\alpha}\lambda_{bi\alpha})$
multiplied by the number of links, it is straightforward to obtain
a threshold value $t_{pp}^{(min)}\approx 0.3t$ above which the
OCP+CDW phase dominates.

 Varma's work was concerned with the strong coupling regime in 2D, and the stability 
of the current patterns was studied in mean-field theory. The fact that we were able to find such a state
in 1D, in the weak coupling limit and with purely repulsive interactions, 
gives an interesting perspective on the possible existence of such orbital
 currents. Note that a type ordering similar to the one we find (current+DW) has been suggested 
in numerical studies of two-leg ladders
\cite{roux_CDW+current, schollwock_CDW+current}. We will return
to this issue in the last part of this section, where we make a
contact with strong coupling results.
The statement about the existence of
OCP states given above, obviously holds also for
the $C2S2$ phase, where CDW fluctuations are replaced by
SDW fluctuations (Fig. \ref{fig:Varma}). We also note two differences between our orbital
current states and Varma's: in our case, the structure is incommensurate (the modulation is doping dependent)
and we  get an additional DW modulation. Hence, the OCP+DW state also share similarities
with the DDW phase \cite{chakravarty_ddw_pseudogap}.
The main point is that introducing $t_{pp}$ gives the
possibility of new types of current loops involving oxygen sites. Phases
with time reversal symmetry breaking have been widely investigated, but,
for single orbital models, currents flow along $Cu$
square plaquettes, giving rise to the OAF state.
As was shown in detail by Fjaerestad and Marston \cite{fjaerestad_staggflux} they are
described by inter-band creation-annihilation process
$\sim\psi_{o}^{\dag}\psi_{\pi}$. 
The order operator in bosonization language is given in Appendix D (it is the
$\hat{O_{OAF}}(r)$ operator) and, this type of quasi-long range order is stable
 \emph{provided} one introduces an attractive $V_{Cu-Cu}$.

What about current patterns in the strong coupling regime? This issue was
 investigated  numerically. One paper
\cite{roux_CDW+current} showed 
that if time reversal symmetry is artificially broken by adding a
magnetic field, one promotes a state with OAF currents and a CDW modulation for the two-leg ladder,
 very similar to the one suggested in Sec. IV.
Another \cite{schollwock_CDW+current} considered variants of t-J models, in hopes of finding
current pattern phases. Although somewhat artificial values were assigned to some of
the parameters, this study suggested that quasi long-range order of the current patterns
could be obtained  provided one changed the internal description of the rung. Furthermore,
the current pattern is accompanied by a charge density wave structure.

It should be pointed out that both papers established a direct connection between
the strong and the weak coupling regimes.
For example, Ref.~\onlinecite{schollwock_CDW+current} showed that the spatial decay of current-current correlations
is similar in regions of parameter space corresponding to weak or large
 coupling RG. This paper also emphasizes that, in order to obtain current patterns, one needs
 to go beyond theories using properties of SO(5) symmetric
models. This confirms our findings, that new physics emerges
in formalisms where symmetry breaking is \emph{a priori} allowed.

\subsection{Experimental systems}

Experimentally, it is rather difficult to vary the doping in ladder compounds,
and the large doping regime is still inaccessible \cite{vuletic_exper_rev}.
Furthermore, different methods 
(NMR, optical conductivity and X-ray measurements)  yield different
values of the doping for a given system \cite{Rusydi_X-ray_ladder1}.  One of the most
interesting compounds, $Sr_{14-x}Ca_xCu_{24}O_{41}$, contains both chains and ladders,
 and it was shown that a change in pressure may cause a charge transfer between
 the two \cite{piskunov_ladder_nmr}. Calcium is also a factor that affects the carrier content of the ladder.
In the low doping regime, this system displays spin-gaps, as
is well established in many NMR studies; this will be discussed in detail in the next section.
As far as charge degree of freedoms are concerned, the situation is more complicated. There are
optical conductivity measurements showing CDW ordering in these systems at ambient
 pressure \cite{gorshunov_CDW_ladder}.
This kind of ordering may be due to a large $V_{Cu-Cu}$ \cite{tsuchiizu_2leg_cdw} not taken
into account in our model or to inter-ladder electrostatic interactions. The
SCd phase appears under pressure, with a maximum temperature of order 10K
for an optimal pressure of 3.5 GPa. The role of pressure in this transition is not clear:
it may change the bandwidth, the couplings between ladders, the
screening of the intra- or inter- ladder interactions, or the doping.
Recently\cite{abbamonte_X-ray_ladder, Rusydi_X-ray_ladder2}, soft X-ray measurements were performed
for this system. Their main conclusion is that an insulating `` hole crystal'' phase
exists for commensurate fillings. It is suggested that this phase melts for
other dopings. The authors interpret their findings by invoking strong on-rung hole
pairing. This analysis supports the picture that emerges from our study of the low doping regime.

\section{NMR properties}

\subsection{Spin susceptibility and NMR relaxation rate}

\quad The spin operator with momentum $q$ is defined as
\begin{equation}\label{spinop}
    S_{m}^{i}(\textbf{q})\equiv
    \frac{1}{2}\sum_{\textbf{k}\sigma1\sigma2}c^{\dag}_{m\sigma1}(\textbf{k}+\textbf{q}) \widehat{\sigma^{i}} c_{m\sigma2}(\textbf{k})
\end{equation}
where $c\equiv a,b$ (respectively the annihilation operator of a hole on $Cu$ or
on $O$) and $\widehat{\sigma^{i}}$ is a
Pauli matrix. From linear response theory, the time-ordered
susceptibility reads
\begin{equation}\label{podat}
    \chi_{m'm}^{i}(\textbf{q},\imath \omega_{n})\equiv
    \frac{1}{2L}\int_{0}^{\beta}d\tau \langle T_{\tau}S_{m'}^{i}(\textbf{q},\tau) S_{m}^{i}(\textbf{-q},0)\rangle
    \exp(\imath \omega_{n} \tau)
\end{equation}
The above function is defined only for Matsubara frequencies
$\omega_{n}$; taking the analytical continuation, one obtains the
retarded susceptibility $\chi_{m'm}^{R}$
\cite{giamarchi_book_1d_ref} and hence
derive \cite{voit_bosonization_revue} analytical expressions for the
measured NMR properties of the system.

The NMR signal comes from a contact interaction between a nucleus and the surrounding cloud of electrons
 in an s-orbital state.

 The temperature dependence of the  shift in
(Zeeman) frequency of the \emph{m-th} nucleus stems from
hoppings of carriers from the m-th atom s-orbital to the highest occupied molecular orbital \emph{p} or
\emph{d} orbital of the neighbouring sites. Thus, the Knight shift is
\begin{equation}\label{eq:Knight}
    \bar{K}^{i}_{m}=\frac{C_{m}}{\gamma_{m}\gamma_{e}\hbar^{2}}\sum_{m'}\chi_{m'm}^{Ri}(\textbf{p}\rightarrow 0, \omega=0)
\end{equation}
where the summation is taken over all neighboring \emph{d-Cu} and
\emph{p-O} orbitals. The overlap coefficients $\widetilde{\lambda_{m\alpha}}$,
which enter  $\chi_{m,m'}$, are evaluated using first
order perturbation theory; we include
hoppings between a \emph{s-Cu} orbital   and
a \emph{p-O} orbital on the neighbor sites or a \emph{d-Cu} orbital on  next-nearest neighbor sites, 
as well as hoppings between a
\emph{s-O} orbital and a \emph{d-Cu} orbital on neighboring sites.

The spin-lattice relaxation rate is also affected by the electronic
environment. The signal measured on the m-th nucleus is given by
\begin{equation}\label{eq:relax}
    (\frac{1}{T_{1m}})^{i}=\frac{C^{2}_{m}}{\gamma_{m}\gamma_{e}\hbar^{2}\beta}\sum_{\textbf{p}}
    \frac{Im[\chi^{Ri}_{mm}(\textbf{q}, \omega_{Zm})]}{\omega_{Zm}}
\end{equation}

In the following, we omit the \emph{`'i"} subscripts, because we
are working with spin-rotationally invariant models. Taking into
account the fact that the Fermi surface consists of pairs of points
of the form $\pm k_{F}$,  the sum in $\frac{1}{T_{1}}$ can be divided into
two independent parts: a uniform piece (\textbf{q} around $k_{\|}=0$) and a
staggered piece (\textbf{q} around $k_{\|}=2k_{F}$).

Using the $\widetilde{\lambda_{m\alpha}}$ allows us to connect the
time-ordered correlation functions $R_{\alpha}(r(x,\tau))$ of
carriers in band $\alpha$  (they are introduced in the bosonic phase
field language), with the $R_{m'm}(r(x,\tau))$ defined for a site basis
\begin{multline}\label{przej}
    R_{m'm}(r(x,\tau))=|\widetilde{\lambda_{m'o}}|^{2}|\widetilde{\lambda_{mo}}|^{2}R_{o}(r(x,\tau))+ \\ |\widetilde{\lambda_{m'\pi}}|^{2}|\widetilde{\lambda_{m\pi}}|^{2}R_{\pi}(r(x,\tau))
\end{multline}
In order to get the retarded $\chi^{R}_{m,m'}(q,\omega)$ entering
Eqs. (\ref{eq:Knight},\ref{eq:relax}),
we use the fact that correlations for spin
operators and for their complex conjugates are equal, and we simply obtain the
retarded spin susceptibility by a Wick rotation
\cite{schulz_spins}:
\begin{equation}\label{retarded}
    \chi_{m'm}^{Ri}(x,t)=2\theta(t)Im[R_{m'm}(r(x,\tau))]_{\tau=\imath t + \delta}
\end{equation}
followed by  Fourier transforming the last function.

Because of conformal symmetry in our 1D quantum theory, results
for zero temperature correlations can be extended to finite
temperatures by simply substituting for the complex coordinates
the following expression
\begin{equation}\label{eq:temp}
    r_{\nu}(x,\tau,\beta)=\frac{u_{\nu}\beta}{\pi}\sqrt{\sinh(\frac{x-\imath u_{\nu} \tau}{\frac{u_{\nu}\beta}{\pi}})
    \sinh(\frac{x+\imath u_{\nu} \tau}{\frac{u_{\nu}\beta}{\pi}})}
\end{equation}
This substitution $ r_{\nu}(x,\tau)\rightarrow
r_{\nu}(x,\tau,\beta)$ gives us the temperature dependence of the
susceptibilities. 

This procedure is valid both for the uniform and for the staggered parts of the
magnetization. We write the
time ordered correlation functions $R_{\alpha}(r(x,\tau))$ in each
band in terms of diagonal modes
$R_{\nu}(r(x,\tau))=F[\phi_{\nu}]$
for the staggered and the uniform part, separately. The form of
$F[\phi_{\nu}]$ depends on whether the $\nu-th$ LL mode is
massless or massive and it will be presented below. Given
$F[\phi_{\nu}]$, the substitution \ref{eq:temp} allows us to obtain
the temperature dependence of $K_{m}$ and $T_{1m}^{-1}$, but as the
temperature increases, the form of $F[\phi_{\nu}]$ is changing.
Generally it is assumed that above the temperature $T_{\nu}$
corresponding to the value of the gap $\Delta_{\nu}$, thermal
fluctuations make the \emph{$\nu$-th} mode massless.
For example, at $T=0$, in the $C1S0$ phase, one starts with three gapped
modes (two for the spin and, one for the charge);  we increase
$T$ until the energy of the first gap $\Delta_{s+}$ is
reached. Above the corresponding temperature $T_{s+}$ we may consider
that there is effectively one gapped and one gapless spin
mode, and similarly for the charge sector. The others gaps ($\Delta_{s-}$
and $\Delta_{c-}$) will successively close at
temperatures $T_{s-}$ and $T_{c-}$.


\subsection{Doping dependence of the NMR signals}
A number of papers
\cite{konik_ff_so8, essler_SO(6)sucs, shelton_spin_ladders},were devoted to the computation of
 magnetic properties of two-leg ladders assuming
symmetry entanglement at the fixed point (\emph{SO(5)} or
\emph{SO(8)}). Yet, following the discussion  in
section IV, we use simpler,
approximate, methods which nevertheless have a wider range of validity.

\subsubsection{Uniform part}

For the uniform magnetization, only spin correlations need
to be taken into account. Because the spin density is
generally related to the spin phase field
$\sigma(x)=\frac{1}{\pi}\nabla~\phi_{\sigma}(x)$, the zero
momentum part of $\bar{R}_{o}$ is a linear combination of bosonic
correlations calculated in the diagonal basis $\bar{R}_{\nu}(r)$.
In the massless case it is known from LL properties and given by
\begin{equation}\label{unifree}
    \bar{R}_{\nu}=\frac{1}{r_{i}^{2}}
\end{equation}
 The contribution to NMR of these power laws has been evaluated many
times before in the literature. One gets a $T^{0}$ dependence
for the Knight shift and $T^{1}$ for the relaxation rate. One
may improve these result both in the high- and low energy limits. At
 low energies, logarithmic corrections from relevant and
marginal couplings $g(\Lambda)$ (where the energy scale $\Lambda$ may
be related to the temperature) should be taken into account. Then,
$g_{1c}$ and $g_{2c}$ contribute to the \emph{o} and $\pi$
bands, $g_{1}+g_{2}$ to the \emph{o} band and $g_{1}+g_{2}$ to the $\pi$
band. At high energies, the curvature of the bands may be
taken into account using an RPA approximation, following Refs.~\cite{tsuchiizu_2leg_NMR, dumoulin96_spinpeierls}.

In the massive case we use the massive Gaussian model to obtain
fluctuations around the static quasi-classical solution
(equilibrium position) and this leads to
\begin{equation}\label{unimass}
     \bar{R}_{\nu}=K_{i}(\kappa_{0}(m_{i}r_{i})+\kappa_{2}(m_{i}r_{i}))
\end{equation}
where we have used the fact that for harmonic fluctuations
around the soliton $\delta\phi_{i}(x)\equiv
\phi_{i}(x)-\phi_{oi}(x)$, correlations are given by a Bessel
function
$\langle\delta\phi_{i}(r)\delta\phi_{i}(0)\rangle\simeq\kappa_{0}(m_{i}r)$.
A first order expansion, valid for large r, gives
exactly the same expression as that found in exact calculations
\cite{essler_rpa_quasi1d, voit_le_spectral}.
One needs to evaluate the following integrals (the exact
formulas for the LL $\chi(k,\omega)$ are known
\cite{iucci_LLsuscFourier} but it is not necessary to use them
here)
\begin{equation}\label{eq:unimass2}
     \frac{1}{T_{1m}}=\int dt \sum_{x_{0}} \digamma[\widetilde{\lambda_{m\nu}}^{2}(x_{0})]R(x_{0},t)
\end{equation}
\begin{equation}\label{eq:unimass32}
     \bar{K}_{m}=\widetilde{\lambda_{m\nu}}^{2}(k=0)(\sum_{m'}\widetilde{\lambda_{m\nu}}^{2}(k=0))\int dt dx R(x,t)
\end{equation}
where the summation over $x_{0}$ accounts for the
momentum dependence of $\widetilde{\lambda_{m\nu}}(k)$. For the uniform part,
integrals can be calculated analytically
\begin{widetext}
\begin{equation}\label{eq:unimass_rez}
    \frac{K_{i}m_{i}V_{i}}{\pi}(\kappa_{0}(m_{i}r_{i})\frac{\cosh(\frac{\pi(x-\imath V_{i} t)}{V_{i}\beta})\sinh(\frac{\pi(x+\imath V_{i} t)}{V_{i}\beta})-\sinh(\frac{\pi(x-\imath V_{i} t)}{V_{i}\beta})\cosh(\frac{\pi(x+\imath V_{i} t)}{V_{i}\beta})}{\sqrt{\sinh(\frac{\pi(x-\imath V_{i} t)}{V_{i}\beta})\sinh(\frac{\pi(x+\imath V_{i}
    t)}{V_{i}\beta})}}\mid_{b}^{c}
\end{equation}
\end{widetext}
the appropriate bound \emph{b,c} is chosen for the Knight shift or for
the relaxation rate, and depends on whether one integrates over a time or a space-time domain.

\begin{figure}[h]
 \centerline{\includegraphics[width=\figwidth]{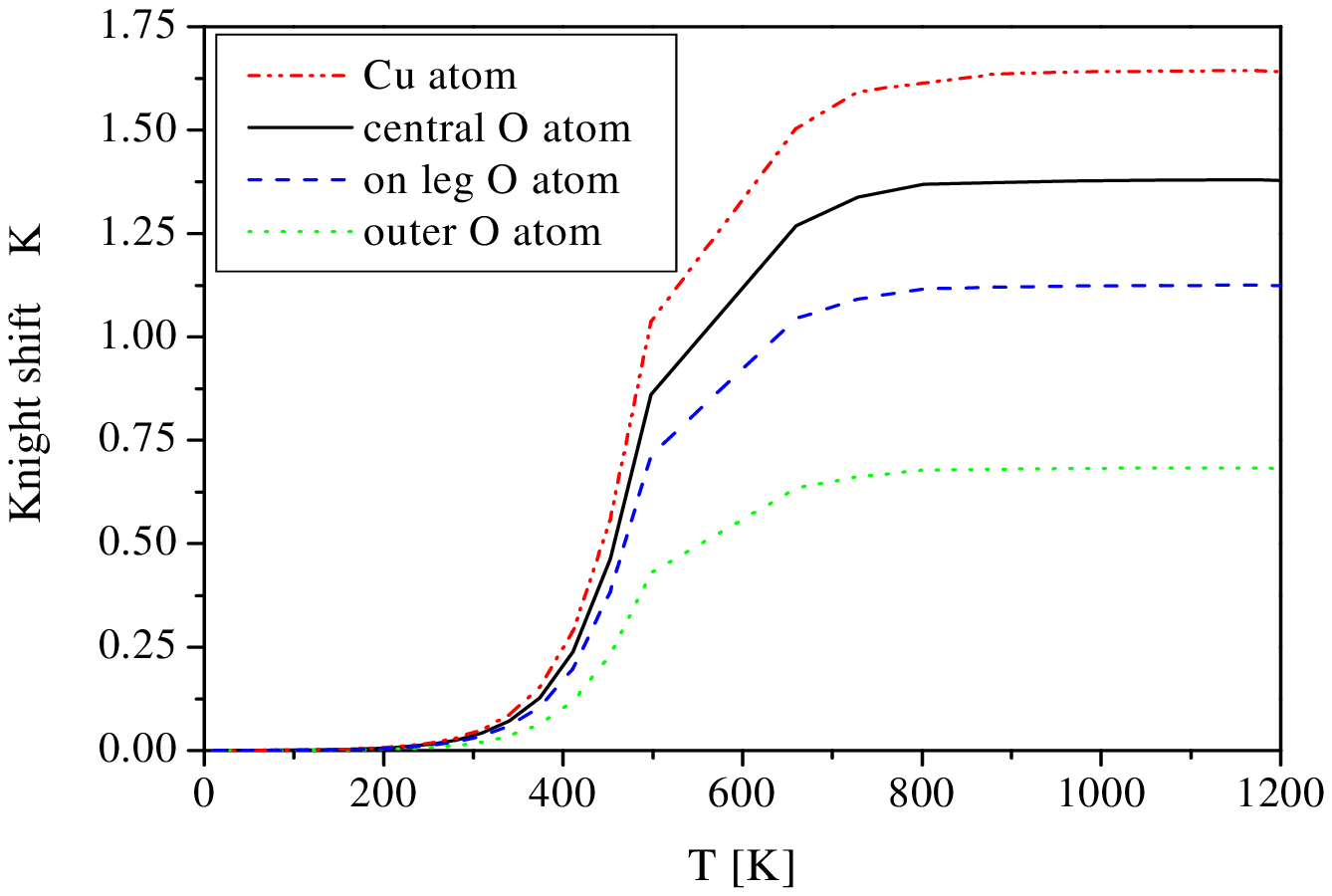}}
  \centerline{\includegraphics[width=\figwidth]{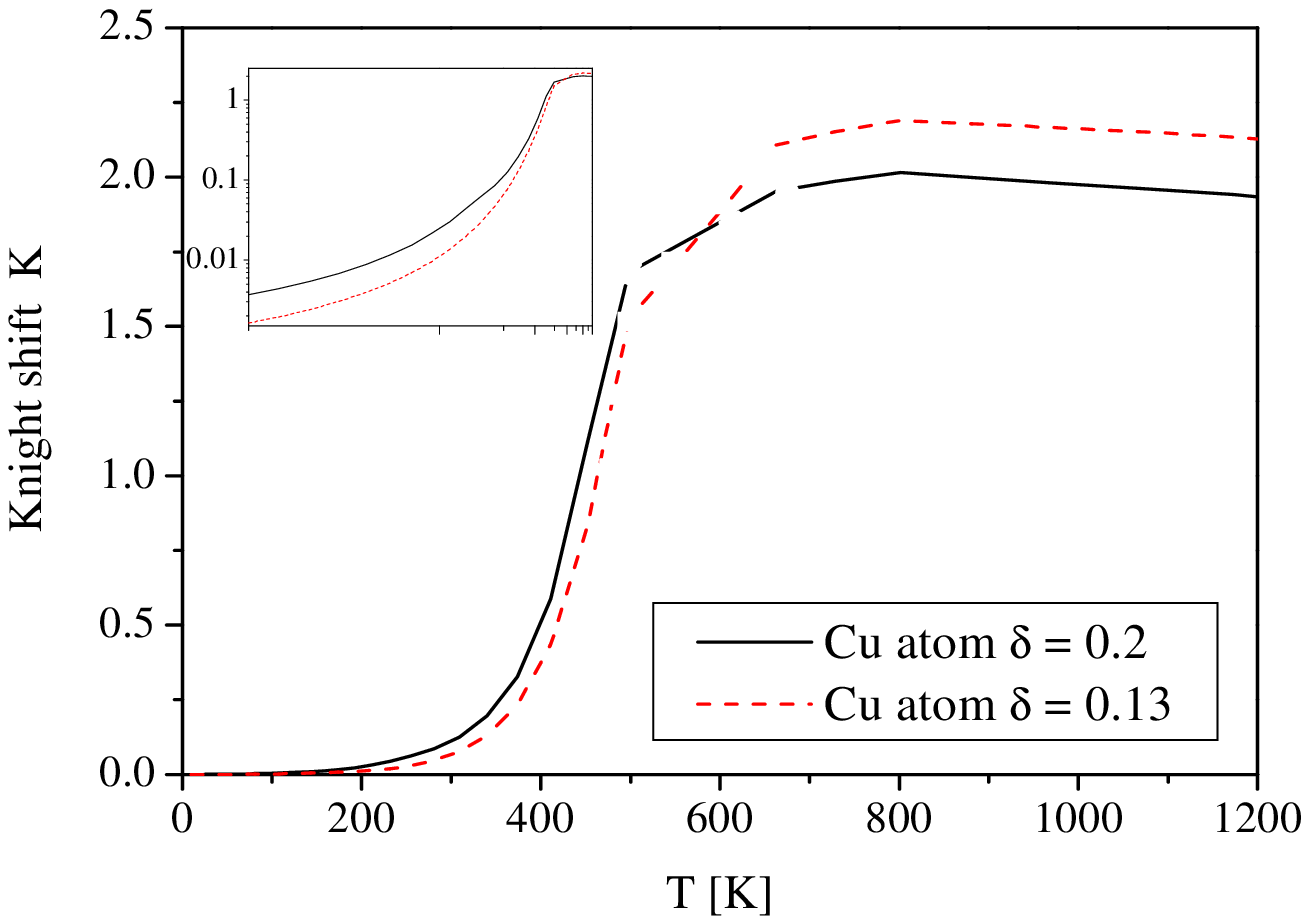}}
  \caption{Temperature dependence of the Knight shifts for
  (a) the different atoms in the elementary cell and
  (b) different dopings in the $C1S0$ phase. The activation gap at low $T$
is shown in the inset}\label{fig:K_C1S0}
\end{figure}

\begin{figure}[h]
  (a)
  \centerline{\includegraphics[width=\figwidth]{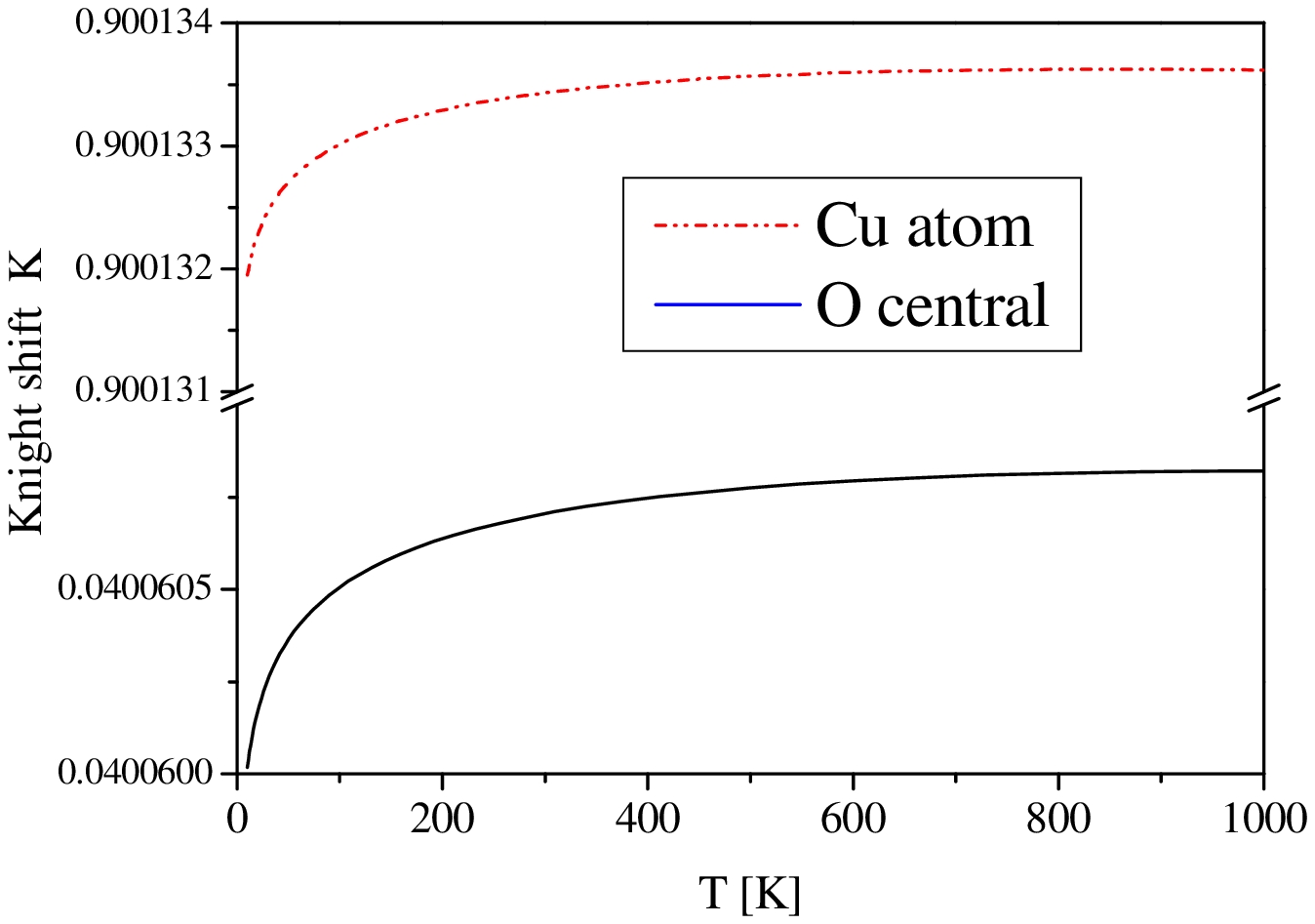}}
  (b)
  \centerline{\includegraphics[width=\figwidth]{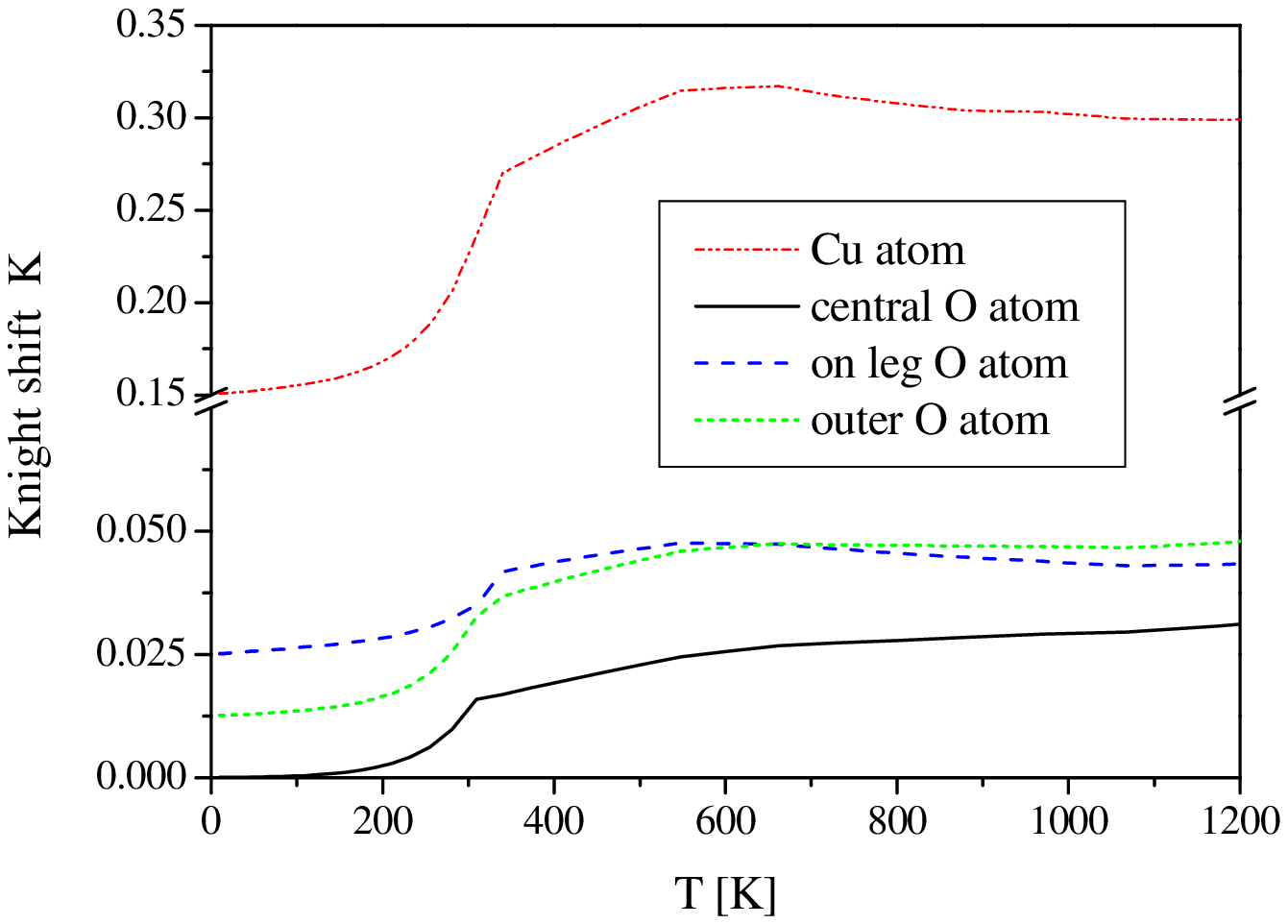}}
  \caption{Temperature dependence of the Knight shifts for
  the different atoms in the elementary cell (a) in the $C2S2$ phase and
  (b) in the $C2S1$ phase}\label{fig:K_C2S1}
\end{figure}

The results for Knight shifts, calculated for different atoms and
different dopings, are shown in Figs. \ref{fig:K_C1S0} and \ref{fig:K_C2S1}. The discussion of
the relaxation rates is postponed until after the evaluation of the staggered part,
  because the quantity which is measured in
experiments is the sum of the uniform and the staggered parts of the
relaxation rate.

For the $C1S0$ phase, an activated behavior
$exp(-\frac{\Delta}{T})$ is seen for the Knight shifts of all the atoms.
 This shows clearly on the logarithmic plots shown in the inset of
Fig. \ref{fig:K_C1S0}. However let us stress that in the
$C1S0$ case we have two spin gaps, so one expects a more complicated
shape than a simple straight line. For higher temperatures the Knight shift
saturates to a constant value. As is expected for the uniform
susceptibility, the responses of the different atoms are similar, only
their amplitudes are different (this is because of the
$|\widetilde{\lambda_{m\nu}}|^{2}$ coefficients). For larger dopings there are
less electrons in the conduction band, and their velocity is
smaller, so the saturation value also decreases. As the doping
increases, spin gaps decrease and 
curves saturate at a lower $T$ until we reach the quantum critical point (QCP)  at $\delta=\delta_{c1}$. 
This behavior for the susceptibility was described in
Ref.~\onlinecite{tsuchiizu_2leg_NMR}. In their case, a QCP appears in the
presence of $V_{Cu-Cu}$; in our case, doping drives the transition.

For the  $C2S1$ phase, we obtain a finite
susceptibility even at zero temperature. This feature comes from
the massless spin mode. The central oxygen atom which is only coupled
to the gapped band does not give a finite susceptibility
at \emph{T=0}. For the second, massive, mode we observe a behavior
similar to the one described above for the $C1S0$ phase, with the
single activation gap shown on the inset of Fig. \ref{fig:K_C1S0}b for two dopings.

For the intermediate doping $C2S2$ phase, including logarithmic corrections is the only
 way to generate some weak $T$ dependence. They arise mainly from the presence of
the marginal $g_{1}$ and $g_{2}$ terms. Their influence on the
uniform susceptibility was described in detail in
Refs.~\cite{tsuchiizu_2leg_NMR, brunel99_edges_logs}.
Differences in the amplitudes of the Knight shifts for the various atoms
in the elementary cell stay pretty much the same from one phase to the next, 
since these amplitudes are simply determined by $\widetilde{\lambda}$ coefficients.

\subsubsection{Staggered part}

\quad For $q=2k_{F}$, both the spin and charge
parts contribute to the band correlation functions. The band
$\tilde{R}_{o/\pi}$ with $2k_{F}$ wave vector is a product of a spin
and and a charge part, $R_{\alpha}=R_{\alpha}^{\sigma}\cdot
R_{\alpha}^{\rho}$. The form of $F[\phi_{\nu}]$
depends on the fixed
point eigenbasis for the angles and on the possible existence of gaps.

The expression for the gapped spin phase was obtained using the
expression for the $2k_{F}$ part of the spin density operator
correlations which is given by $\langle O_{o/\pi SDW}(r)O_{o/\pi
SDW}(0)\rangle^{2k_{F}} \sim \cos(\phi_{1}\pm\phi_{2})$. The last form
could be evaluated using the fact that
$\phi_{i}=\phi_{oi}+\delta\phi_{i}$ where the fluctuations of
$\delta\phi_{i}$ are described by a massive Gaussian model, as
was shown in the case of the uniform part. For gapped spin $(\sigma)$ modes
there are two possibilities
\begin{multline}\label{eq:stagmass}
    \tilde{R}_{o/\pi}^{\sigma}=\sinh(K_{2}\kappa_{0}(m_{2}r_{2})\pm
    K_{1}\kappa_{0}(m_{1}r_{1}))(m_{2}a)^{K_{2}}(m_{1}a)^{K_{1}}\\
    \textrm{for the \emph{C1S0} phase}\\
    \tilde{R}_{o}^{\sigma}=\sinh(K_{2}\kappa_{0}(m_{2}r_{2}))(m_{2}a)^{K_{2}}\\
    \textrm{for the \emph{C2S1} phase}
\end{multline}

In the gapless case, one gets a power law behavior; for the high-$T$ limit
of the $C1S0$ phase, where the $B_{+-}$ eigenbasis  is relevant, we find
\begin{equation}\label{stagfree1}
\tilde{R}_{\alpha}=(
\frac{1}{r_{i}})^{K_{i}/2}(\frac{1}{r_{j}})^{K_{j}/2}
\end{equation}
where $\alpha$ corresponds to the band index and $i=1,3$, $j=2,4$ are
the LL modes;

For the $C1S0$ phase, the charge mode is only partially gapped: the field
$\theta_{3}$ is locked so the charge antisymmetric mode does not
give any contribution to SDW, but the massless ,``4'' (charge symmetric)
mode gives a power law contribution
\begin{equation}\label{stagfree}
    \tilde{R}_{o/\pi}^{\rho}=(\frac{1}{r})^{K_{4}/2}
\end{equation}
For the other phases, both charge modes are massless and in this case,
 $B_{o\pi}$ is the fixed point basis, and we have
\begin{equation}\label{stagfree3}
 \tilde{R}_{o}=(\frac{1}{r_{j}})^{K_{4}}
\end{equation}
\begin{equation}\label{stagfree2}
 \tilde{R}_{\pi}=(\frac{1}{r_{i}})^{K_{3}}
\end{equation}
For the spin part in the $C2S1$ phase, one substitutes $K_{1},K_{2}$
to $K_{3},K_{4}$.
The amplitudes of $\tilde{R}_{\alpha}$ on different atoms
need to be calculated. Once again $\widetilde{\lambda_{m\alpha}}(k)$ are
involved, however for those atoms with neighbors along the
ladder (on-leg $Cu$ and $O$ atoms) these coefficients are
different, because of phase factors at $k=2k_{F}$ which cause cancellations
in some contributions of neighboring atoms.
Another possible factor may cause  differences between atoms
in the elementary cell. Following
Ref.~\onlinecite{becker_NMR}, one may assume that, below a characteristic distance
$x<L_{\delta}=\delta^{-1}$, umklapp terms are relevant and
that they open up a gap in the charge symmetric channel. This massive charge
correlation affects the staggered part of the magnetic
susceptibility, and yields an expression similar to
Eq. (\ref{eq:stagmass}) (with $\cosh$ instead of $\sinh$). For
on-leg oxygens, which sit between two $Cu$ along the ladder, one recovers
a $\sinh$ instead of a $\cosh$. This produces different amplitudes
$\tilde{R}_{o/\pi}^{\rho}(L)$ for $Cu$ and for $O$ on-leg atoms,
provided
$L_{\delta}>m_{4}^{-1}$. We have made the calculation for the half
filled case, and the result is that $m_{4}$ is of the same order
as $m_{2}$, so for dopings larger than $0.05$ this effect
should not play any role.

Once band correlation functions are known one may follow
exactly the same procedure as in the uniform case in order to
obtain the temperature dependence of $\frac{1}{T_{1}}$.


\subsubsection{Total relaxation rate}

\begin{figure}[h]
%
%
  \centerline{\includegraphics[width=\figwidth]{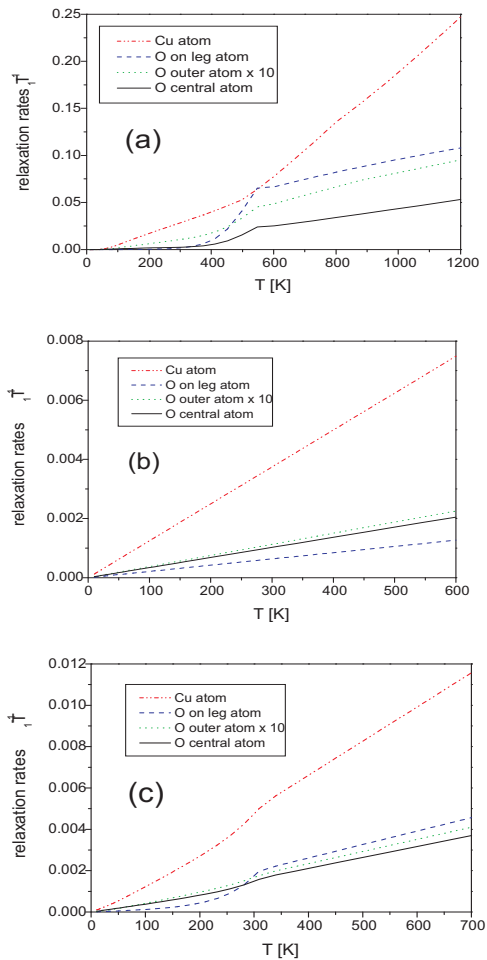}}
  \caption{Temperature dependence of the relaxation rates for
  the different atoms in the elementary cell (a) in the $C1S0$ phase,(b) in the $C2S2$ phase,
  (c) in the $C2S1$ phase}\label{fig:T1}
\end{figure}

The plots in Fig . \ref{fig:T1} shows $1/T_{1}$ for different atoms in the elementary
cell. They were obtained by numerical integration of
Eq. (\ref{eq:stagmass}) and adding the result to that computed for the
uniform part.

As for the Knight shifts, the difference between atoms are
caused mainly by the different $\widetilde{\lambda_{m\alpha}}$ coefficients.
However these coefficients can be different for the staggered part
and for the uniform part.

The first observation is the linear dependence of
$\frac{1}{T_{1}}$ at high temperatures for all atoms, for all
dopings. The Knight shift saturates in this temperature range to a
constant value and this is in accordance with the Korringa law. 
 For the $C2S2$ phase we observe the
linear dependence as expected for a massless LL with all
\emph{K} parameters close to one.

%
%
The second main conclusion is that processes involving large
$k_{\parallel}$ transfers can strongly affect the measured rates,
especially for temperatures comparable with the spin gaps,
as previously reported \cite{ivanov_nmr_ladder_stag}. One
observes only small differences between atoms in the elementary cell
at low $T$. The difference in the relaxation rate of a $Cu$ nucleus compared with a
central $O$ nucleus comes form the fact that the latter may only relax
through processes in the ``$\pi$'' band, while, for the
former, both bands contribute. The difference between on-leg
atoms and atoms sitting at other locations comes from the fact that
the staggered part contribution of the former nucleus is very small, as it is
suppressed by the opposite contributions of the two neighboring $Cu$ atoms. The low-$T$ activation behavior (in the $C2S1$ and $C1S0$
phases) is then clearly seen on these on-leg $O$ sites.

Two points should be kept in mind when comparing our
results with experiments. First our $\Lambda_{0}$ is of order
\emph{0.5 eV}, so that the largest charge antisymmetric gap is of order
 \emph{700K}; observing it would be 
experimentally challenging, and it would be even harder to reach the Korringa regime predicted at higher $T$.
Second, our calculations were made in the phase where
SCd fluctuations dominate. Thus experiments done at
large pressures would be the most relevant to compare
our findings with.

Aside from the above caveats, the results of our
calculations seem to be in very reasonable agreement with experiments
\cite{fujiwara03_ladder_supra, piskunov04_sr14cu24o41_nmr}.

\section{Conclusions}

Our study has clearly shown that including oxygen
atoms in the structure produces  significant changes in
the ground state phase diagram of doped, $Cu$-$O$, two-leg Hubbard ladders.
This result is fully consistent with DMRG studies suggesting that there are quantitative differences
between  models which include $O$ atoms and models which do not, even close to half-filling.
The massless $C2S2$ phase is of special importance in
that respect. A \emph{Varma-like} phase with incommensurate orbital current patterns
and additional density wave characterize the ground state structure
at intermediate and large dopings. Signatures of these states can be seen in NMR experiments 
probing the various nuclei in the cell.

We see important differences between $Cu$ and $Cu$-$O$ ladders in the weak interaction limit ($U<\epsilon$), but numerical approaches which can investigate the opposite limit as well suggest 
that these differences do survive for $U>\epsilon$.
This invites further analytical studies
of the two-leg $Cu$-$O$ Hubbard ladders in the large $U$ limit.


\begin{acknowledgements}
This work was supported in part by the Swiss NSF under MaNEP and Division II and by an ESRT Marie Curie fellowship.
\end{acknowledgements}

\appendix

\section{Coupling constants} \label{ap:couplings}

Initial conditions for the non-linear terms are
\begin{equation}
\begin{split}
 g_{1c}^{o} &=4f(1,2,1,2,1) \\
 g_{1a}^{o} &=4f(1,2,2,1,1) \\
 g_{2c}^{o} &=4f(1,2,1,2,-1) \\
 g_{\parallel c}^{o} &=4(f(1,2,1,2,1)-f(1,2,1,2,-1)) \\
 g_{1}^{o} &=\frac{2f(1,1,1,1,1)}{V_{Fo}}+\frac{2f(2,2,2,2,1)}{V_{F\pi}} \\
 g_{2}^{o} &=\frac{2f(1,1,1,1,1)}{V_{Fo}}-\frac{2f(2,2,2,2,1)}{V_{F\pi}} \\
 g_{4a}^{o} &=4f(1,2,2,1,-1)
\end{split}
\end{equation}
where the function $f(k,l,m,n,p)$ converts the interactions
given in the atomic basis ($U_{Cu},U_{O},V_{Cu-O}$) into band ``g-ology'' interactions.
\begin{multline}\label{transl}
    f(k,l,m,n,p)=\sum_{i,j}\lambda_{ik}^{\ast}\lambda_{il}^{\ast}\lambda_{jm}\lambda_{jn}(V^{intra}+ \\
    V^{inter}\cos(k_{Fm}-p\cdot k_{Fn}))
\end{multline}
The summation is taken over all the atoms in the elementary
cell. $V^{intra}$ denotes interactions within the elementary
cell, and $V^{inter}$ is $V_{Cu-O}$, since one of the
atoms is outside the elementary cell, as in
Ref.~\onlinecite{lee_marston_CuO}. Initial values for $K$ and
$\cot(\alpha), \cot(\beta)$ are evaluated as follows: starting from Eq. (\ref{eq:matrix}),
one performs a $S(\pi/4)$ rotation. In this
$B_{+-}$ basis, $\hat{K}$ can be calculated by
simply solving a matrix equation. The initial $K_{\nu}$ are given
by the eigenvalues of this matrix, and $\cot(\alpha), \cot(\beta)$ are
the ratios of non-diagonal terms to the difference of diagonal
ones. For example: $\cot(\alpha)=2B_{s+,s-}/(K_{s-}-K_{s+})$.

The system of RG differential equations is solved by means of an iterative
method.

If
$\cot(\alpha)$ (or$\cot(\beta)$) becomes very large during the flow, we stop
the flow at some point, introduce the tangent of the angles instead of the cotangent,
and then resume the iteration scheme.
In this way, we are able to isolate divergences
of the prefactors in some of the cosine terms, which cause gaps to open and affect observables.

\section{Derivation of the RG equations} \label{ap:RG}

\subsection{Flow of the diagonal basis}

To second order in perturbation, one finds the corrections  $dK_{1}$, $dK_{2}$, $dK_{3}$, $dK_{4}$
 to the LL parameters, and  the non-diagonal terms $dB_{12}$, $dB_{34}$.
These non-diagonal terms signal that, after the RG step, $B_o$ is no longer
a diagonal basis. We then go back to
the $B_{+-}$ basis, using the transformation $ S^{-1}$.
In this basis, off-diagonal terms have been incremented by small amounts during the RG step.
For instance
\begin{equation}
 dB_{s-s+}= -\frac{1}{2}(dK_{1}-dK_{2})\cos2\alpha +
 dB_{12}\sin2\alpha
\end{equation}

Diagonal terms also undergo infinitesimal variations
\begin{equation}
 dK_{s-(s+)}=\frac{1}{2}(dK_{1}+dK_{2})\pm
 \frac{1}{2}(2dB_{12}\cos2\alpha+(dK_{1}-dK_{2})\sin2\alpha)
\end{equation}

Similar expressions hold for the charge modes when we perform the substitutions $s\to c$, $1\to 3$, $2\to 4$,
 $\alpha \to \beta$ in the equations above.

The new matrix is diagonalized by the operator $S(\alpha+d\alpha,\beta+d\beta)$, where the angle $d\alpha$,
 which account for
the  $ dB_{\mu-\mu+} $ and $ dK_{\mu-(\mu+)}$ variations ($\mu=c,s$), indicate a rotation of $B_o$.
 This idea is summarized in the diagram shown in \fref{fig:obrot}.

We now determine the renormalization flow of the angles $ \alpha $ and $\beta $.
In the spin sector, the diagonalization condition 
is written in terms of $ K_{s-(s+)} $ and $ B_{s-s+}$
\begin{equation}
 \frac{1}{2}(K_{s-}-K_{s+})\cos2\alpha + B_{s-s+}\sin2\alpha=0
\end{equation}
One differentiates the above equation in order to relate
 $d\alpha$, $ K_{s-(s+)} $ and $ B_{s-s+}$:
\begin{multline}
 -\frac{d2\alpha}{\sin^{2}2\alpha}=\frac{2}{K_{s-}-K_{s+}}\cdot\\
 \frac{\frac{1}{2}(dK_{s-}-dK_{s+})\cos2\alpha-dB_{s-s+}\sin2\alpha}{(\sin^{2}2\alpha-\cos^{2}2\alpha)\sin2\alpha}
\end{multline}
In the diagonal basis this equation reads
\begin{multline}
 d\cot2\alpha=-\frac{1}{K_{1}-K_{2}}((dK_{1}-dK_{2})\frac{2\sin2\alpha\cos2\alpha}{\sin^{2}2\alpha-\cos^{2}2\alpha}-\\
 dB_{12}\frac{-\sin^{2}2\alpha+\cos^{2}2\alpha}{\sin^{2}2\alpha-\cos^{2}2\alpha})
\end{multline}
where the differentials of the LL parameters are known in the
diagonal basis. They were obtained to second order in
perturbation, and the $dK_{\nu}$, which we use here, were given in
\sref{ICfilRG}. In the charge sector, we obtain the equivalent set of equations
with the changes $s\to c$, $1\to 3$, $2\to 4$, $\alpha\to \beta$. 
 The additional expressions for the differentials of off-diagonal terms
are obtained in a similar way, giving, for the case of a generic filling
\begin{equation}
\begin{split}
 dB_{12} &=P_{1}Q_{1}((g_{1a}^{2}+g_{\parallel c}^{2}+G_{t}^{2})- \\
  & K_{1}K_{2}(g_{1a}^{2}+g_{1c}^{2}+g_{2c}^{2}+G_{p}^{2}))-K_{1}K_{2}h(P_{1})g_{1}g_{2} \\
 dB_{34} &=P_{2}Q_{2}(g_{1c}^{2}+g_{2c}^{2}+ g_{\parallel c}^{2})
\end{split}
\end{equation}
where $h(P_{1})=((P_{1}Q_{1})^{2}+0.25(P_{1}^{2}-Q_{1}^{2}))^{-1}$.

\subsection{First order correction to $g_{1}$, $g_2$}

Setting $g_{1}=g_{1d}+g_{1d'}$ and
$g_{2}=g_{1d}-g_{1d'}$,  the Hamiltonian reads
\begin{multline}
 H= H_{LL} + g_{1}\int dr \cos(2\phi_{s-})\cos(2\phi_{s+})+\\
 g_{2}\int dr \sin(2\phi_{s-})\sin(2\phi_{s+})
\end{multline}
In order to simplify the RG calculation,  we first solve this
problem in the total/transverse basis  where the
averages over the high energy terms are $\langle
\phi_{s-}(r)^{2}\rangle_{h}=K_{s-} dl $, $ \langle
\phi_{s+}(r)^{2}\rangle_{h} =K_{s+}dl $ and $ \langle
\phi_{s-}(r)\phi_{s+}(r)\rangle_{h} = \frac{1}{2} A dl $. One may
determine the renormalization flow that is produced when integrating out the
high energy components; for instance, the renormalization of $g_{1}$ gives
\begin{multline}
 \langle g_{1} \int dr \cos(2(\phi_{s+}+h[\phi_{s+}]))\cdot
 \cos(2(\phi_{s-}+h[\phi_{s-}]))\rangle_{h} = \\
 \frac{1}{2} g_{1} \langle \int dr
 \cos(2(\phi_{s+}+\phi_{s-})+(h[\phi_{s+}]+h[\phi_{s-}]))\rangle_{h}+\\
 + \langle \int dr
 \cos(2(\phi_{s+}-\phi_{s-})+(h[\phi_{s+}]-h[\phi_{s-}]))\rangle_{h}
\end{multline}
We reexponentiate the cosines, use Debye-Waller type relations and
expand the exponential function in Taylor series
\begin{multline}
    \langle \cos(x+h[x]) \rangle_{h} = \cos(x)\langle
    \sum_{\sigma=\pm}\exp(2 \imath \sigma h[x])\rangle_{h}\\
    = \cos(x)\exp (- \langle h[x]^{2} \rangle_{h}) = (1-\langle
    h[x]^{2} \rangle_{h}) cos(x)
\end{multline}
where $\langle
(h[\phi_{s+}]\pm h[\phi_{s-}]))^{2} \rangle_{h}= (K_{s+}+K_{s-})dl
\pm Adl $. One then finds the usual diagonal term
\begin{multline}
 g_{1} (1-(K_{s+}+K_{s-})dl)\int dr ( \cos(2(\phi_{s+}+\phi_{s-}))\\
 +\cos(2(\phi_{s+}-\phi_{s-}))=\\ g_{1} (1-(K_{s+}+K_{s-})dl)\times\\
 \times\int dr \cos(2 \phi_{s+})\cos(2\phi_{s-})
\end{multline}
After rescaling the integration variable dr one gets the
RG equation for $ g_{1} $. But in the non-diagonal basis there is
also an additional term
\begin{multline}
 g_{1} A dl \int dr ( \cos(2(\phi_{s+}+\phi_{s-}))-
 \cos(2(\phi_{s+}-\phi_{s-}))=\\ g_{1} A dl \int dr
 \sin(2\phi_{s-})\sin(2\phi_{s+})
\end{multline}
This links the change of $ g_{2} $  to the
coupling constant $ g_{1}$. The derivation of the RG equation for the $g_2$ term
is obtained in a similar fashion, using the identity:
\begin{equation}
 \sin(2\phi_{s+})\sin(2\phi_{s-})=\frac{1}{2}(\cos(\phi_{s+}-\phi_{s-})-\cos(\phi_{s+}+\phi_{s-}))
\end{equation}
Finally, the first-order RG equation for $ g_{1}$ is
\begin{equation}
\frac{dg_{1(2)}}{dl} = g_{1(2)}(2-(K_{s+}+K_{s-}))+ g_{2(1)} A
\end{equation}
In the diagonal basis, using
\begin{equation}
\begin{split}
 K_{s-(2)} & \rightarrow P^{2}K_{1(2)}+2PQ A_{12} + Q^{2}K_{2(1)}\\
 A & \rightarrow PQ (K_{1}-K_{2})+ A_{12}(P^{2}-Q^{2})
\end{split}
\end{equation}
The RG equations for the couplings are
\begin{equation}
\frac{dg_{1(2)}}{dl}=g_{1(2)}(K_{2}+K_{1}) +
g_{2(1)}((P^{2}-Q^{2})A_{12}+PQ (K_{1}-K_{2}))
\end{equation}

\subsection{RG equations for the half filled case} \label{ap:comm}

Using the same method as for the incommensurate case we find the
following system of equations
\begin{widetext}
\begin{equation}\label{renormset}
\begin{split}
\frac{dK_{1}}{dl} &=\frac{1}{2}[P_{1}^{2}(g_{1a}^{2}+
J_{0}(\delta)g_{3a}^{2}+g_{\parallel
c}^{2}+G_{t}^{2})-K^{2}_{1}(Q_{1}^{2}g_{1a}^{2}+J_{0}(\delta)Q_{1}^{2}g_{3\parallel}^{2}+
Q_{1}^{2}g_{1c}^{2}+P_{1}^{2}G_{p}^{2}+
P_{1}^{2}g_{2c}^{2}+J_{0}(\delta)P_{1}^{2}g_{3b}^{2}\\ & +\frac{1}{2}(g_{1}^{2}+g_{2}^{2})+f(P_{1})
(g_{1}g_{2}))]\\
\frac{dK_{2}}{dl} &=\frac{1}{2}[Q_{1}^{2}(g_{1a}^{2}+
J_{0}(\delta)g_{3a}^{2}+ g_{\parallel
c}^{2}+G_{t}^{2})-K^{2}_{2}(P_{1}^{2}g_{1a}^{2}+P_{1}^{2}g_{1c}^{2}+J_{0}(\delta)P_{1}^{2}g_{3\parallel}^{2}
+Q_{1}^{2}G_{p}^{2}+J_{0}(\delta)Q_{1}^{2}g_{3b}^{2}+
Q_{1}^{2}g_{2c}^{2}\\ & +\frac{1}{2}(g_{1}^{2}+g_{2}^{2})-f(P_{1})(g_{1}g_{2}))]\\
\frac{dK_{3}}{dl} &=\frac{1}{2}P_{2}^{2}[g_{1c}^{2}+g_{2c}^{2}+
g_{\parallel
c}^{2}+g_{3c}^{2}]+\frac{1}{2}Q_{2}^{2}(g_{3\parallel}^{2}+g_{3a}^{2}+g_{3b}^{2}+g_{3c}^{2})\cdot
    J_{0}(\delta)\\
\frac{dK_{4}}{dl} &=\frac{1}{2}Q_{2}^{2}[g_{1c}^{2}+g_{2c}^{2}+
g_{\parallel
c}^{2}+g_{3c}^{2}]+\frac{1}{2}P_{2}^{2}(g_{3\parallel}^{2}+g_{3a}^{2}+g_{3b}^{2}+g_{3c}^{2})\cdot
    J_{0}(\delta)\\
\frac{dg_{1c}}{dl} &=g_{1c}[2-(P_{1}^{2}K_{2}+P_{2}^{2}K_{3}^{-1}+Q_{1}^{2}K_{1}+Q_{2}^{2}K_{4}^{-1})]-
(g_{1}g_{2c}+g_{1a}g_{\parallel
c}+J_{0}(\delta)g_{3c}g_{3\parallel})\\
\frac{dg_{1a}}{dl} &=g_{1a}[2-(P_{1}^{2}(K_{2}+K_{1}^{-1})+Q_{1}^{2}(K_{1}+K_{2}^{-1}))]-
(g_{1c}g_{\parallel c}+J_{0}(\delta)g_{3a}g_{3\parallel})\\
\frac{dg_{2c}}{dl} &=g_{2c}[2-(P^{2}(P_{2}^{2}K_{3}^{-1}+P_{1}^{2}K_{1}+Q_{2}^{2}K_{4}^{-1}+Q_{1}^{2}K_{2})]-(g_{1c}g_{1}+J_{0}(\delta)g_{3c}g_{3a})\\
\frac{dg_{1}}{dl} &=g_{1}(2-(K_{2}+K_{1}))+
P_{1}Q_{1}(K_{2}-K_{1})g_{2} -\gamma
(g_{1c}g_{2c}+J_{0}(\delta)g_{3b}g_{3\parallel})\\
\frac{dg_{2}}{dl} &=g_{2} (2-(K_{2}+K_{1}))+
P_{1}Q_{1}(K_{2}-K_{1})g_{1}\\
\frac{dg_{\parallel c}}{dl} &=g_{\parallel c}
(2-(P_{1}^{2}K_{1}^{-1}+Q_{1}^{2}K_{2}^{-1}+P_{2}^{2}K_{3}^{-1}+Q_{2}^{2}K_{4}^{-1}))
-(g_{1a}g_{1c}+J_{0}(\delta)g_{3a}g_{3c})\\
\frac{dg_{4a}}{dl} &=
g_{4a}(2-\frac{1}{2}(P_{1}^{2}(K_{1}+K_{1}^{-1})+Q_{1}^{2}(K_{2}+K_{2}^{-1})))\\
\frac{dG_{p}}{dl} &=G_{p}(1-(P_{1}^{2}K_{1}+Q_{1}^{2}K_{2}))+g_{4a}^{2}(P_{1}^{2}(K_{1}-K_{1}^{-1})+Q_{1}^{2}(K_{2}-K_{2}^{-1}))\\
\frac{dG_{t}}{dl} &=G_{t}(1-(P_{1}^{2}K_{1}^{-1}+Q_{1}^{2}K_{2}^{-1}))+g_{4a}^{2}(P_{1}^{2}(-K_{1}+K_{1}^{-1})+Q_{1}^{2}(-K_{2}+K_{2}^{-1}))\\
\frac{dg_{3\parallel}}{dl} &=g_{3\parallel}
(2-(P_{1}^{2}K_{2}+P_{2}^{2}K_{4}+Q_{1}^{2}K_{1}+Q_{2}^{2}K_{3}))-(g_{1}g_{3b}+g_{1c}g_{3c}+g_{1a}g_{3a})\\
\frac{dg_{3a}}{dl} &=g_{3a}
(2-(P_{1}^{2}K_{1}^{-1}+P_{2}^{2}K_{4}+Q_{1}^{2}K_{2}^{-1}+Q_{2}^{2}K_{3}))
-(g_{\parallel c}g_{3c}+g_{1a}g_{3\parallel})\\
\frac{dg_{3b}}{dl} &=g_{3b}
(2-(P_{1}^{2}K_{1}+P_{2}^{2}K_{4}+Q_{1}^{2}K_{2}+Q_{2}^{2}K_{3}))
-(g_{1}g_{3\parallel}+g_{2c}g_{3c})\\
\frac{dg_{3c}}{dl} &=g_{3c}
(2-(P_{2}^{2}(K_{4}+K_{3}^{-1})+Q_{2}^{2}(K_{3}+K_{4}^{-1})))
-(g_{\parallel c}g_{3a}+g_{2c}g_{3b}+g_{1c}g_{3\parallel})
\end{split}
\end{equation}
\end{widetext}
where
\begin{equation}
 f(P_{1})=(P_{1}Q_{1}+\frac{1}{4}\frac{P_{1}^{2}-Q_{1}^{2}}{P_{1}Q_{1}})^{-1}
\end{equation}

The renormalization of the parameter $\gamma$ is controlled  by the same
equation as before. The additional flows for the velocities of the
modes, due to umklapp scattering, are all proportional to a Bessel term
$J_{2}(4\delta)$, and hence neglected. The general formula describing the
flow of the diagonal basis remains the same as for the incommensurate case,
but one needs to substitute modified expressions of the $dK_{\nu}$.

\section{Order parameter operators in bosonization language} \label{ap:operators}

We first write the order parameters in fermionic language.
We only consider those order parameters which can produce power-law decays of correlations for the various
locked phase fields combinations. These operators are first defined
for each site, then expressed in the $o/\pi$ basis where
the $\lambda_{m\alpha}$ coefficients enter their expressions.

There are two kinds of order parameter operators. The first group represent charge
density (particle-hole) fluctuations with a $2~k_{F}$ wave
vector. They correspond to the usual CDW, which is the sum of CDW in each
band. Up to an unimportant constant factor it gives
\begin{equation}
 O_{CDW}\sim\sum_{\mu}\sum_{\sigma\sigma'}
 \alpha^{\dag}_{-\mu\sigma}\delta_{\sigma\sigma'}\alpha_{+\mu\sigma'}
\end{equation}
The subscript $\mu$ denotes the band, so, to obtain
the order parameter inside one specific band it is enough to take
the first or the second term in the above sum. It is also possible
to define an operator which describes the difference of the
densities on the two legs
\begin{equation}
O_{\pi CDW}\sim\sum_{\mu}\sum_{\sigma\sigma'}
\alpha^{\dag}_{-\bar{\mu}\sigma}\delta_{\sigma\sigma'}\alpha_{+\mu\sigma'}
\end{equation}
or the operator which describes an orbital antiferromagnetic (OAF) fluctuation where currents
flow along the legs and the rungs of the ladder
\begin{equation}
O_{OAF}\sim\sum_{\mu}\sum_{\sigma\sigma'}
\mu\cdot\alpha^{\dag}_{-\bar{\mu}\sigma}\delta_{\sigma\sigma'}\alpha_{+\mu\sigma'}
\end{equation}

We can also define (at half filling) "bond" operators, which represent density waves
located on the bonds, either in phase
\begin{equation}
O_{BDW}\sim\sum_{\mu}\sum_{\sigma\sigma'}
\exp(k_{F\mu}x)\alpha^{\dag}_{-\mu\sigma}\delta_{\sigma\sigma'}\alpha_{+\mu\sigma'}
\end{equation}
out of phase between the two legs of the ladder,
\begin{equation}
O_{\pi BDW}\sim\sum_{\mu}\sum_{\sigma\sigma'}
\exp(k_{F\mu}x)\alpha^{\dag}_{-\bar{\mu}\sigma}\delta_{\sigma\sigma'}\alpha_{+\mu\sigma'}
\end{equation}
or in the diagonal direction:
\begin{equation}
O_{FDW}\sim\sum_{\mu}\sum_{\sigma\sigma'}
\exp(k_{F\mu}x)\mu\cdot\alpha^{\dag}_{-\bar{\mu}\sigma}\delta_{\sigma\sigma'}\alpha_{+\mu\sigma'}
\end{equation}
Away from half-filling, on-site and bond operators are degenerate, because of
translational invariance (the charge symmetric mode is masless).

The second group describes superconducting pairing (particle-particle) fluctuations with zero wave vectors.
 As usual there is
the  $s$-wave pairing
\begin{equation}
 O_{SCs}\sim\sum_{\mu}\sum_{\sigma\sigma'}
 \sigma\alpha_{-\bar{\mu}\bar{\sigma}}\delta_{\sigma\sigma'}\alpha_{+\mu\sigma'}
\end{equation}
and the $d$-wave pairing, which corresponds to a change of sign of the order parameter
 when moving from along the
legs to along the rungs
\begin{equation}
O_{SCd}\sim\sum_{\mu}\sum_{\sigma\sigma'}
\sigma\mu\alpha_{-\mu\bar{\sigma}}\delta_{\sigma\sigma'}\alpha_{+\mu\sigma'}
\end{equation}

These phases are given different names in the literature.
The name orbital antiferromagnet (OAF) was
used traditionally for the operator defined above, but it is
also called staggered flux\cite{tsuchiizu_2leg_firstorder}
(SF) or d-density wave\cite{wu_2leg_firstorder} phase (DDW).
Its bond counterpart is sometimes called  f-density wave\cite{tsuchiizu_2leg_firstorder} (FDW), or
diagonal current\cite{wu_2leg_firstorder} (DC) phase. Similarly, our $\pi CDW$ and
$\pi BDW$ orders,  are also denoted\cite{tsuchiizu_2leg_firstorder} CDW and PDW or CDW and SP
in Ref.~\onlinecite{wu_2leg_firstorder}. We have decided to use the notation $\pi
CDW$ to avoid any confusion with the usual CDW which also appears in our calculation.

We can now represent the operators in terms of boson fields, using the mapping
(\ref{eq:mapbos}). It is important to keep the same convention for the signs of the Klein factors as
that we used to write the Hamiltonian in bosonic form. Choosing
$\Gamma=+1$ we get $\eta_{\sigma+}\eta_{\sigma-}=+\imath$. This
determines whether a $\sin$ or a $\cos$ appears in the formulas
below.
This choice was used in Ref.~\onlinecite{tsuchiizu_2leg_firstorder} and Ref.~\onlinecite{lee_marston_CuO}
but the opposite
one was used in Ref.~\onlinecite{becker_NMR}. One can easily relate the two by shifting 
the phase fields $\phi$ by an amount $\pi/2$.

The operators take the form
\begin{widetext}
\begin{equation}
\begin{split}
 O_{\pi CDW} &\sim\cos\phi_{c+}\sin\theta_{c-}\cos\phi_{s+}\cos\theta_{s-}-
 \sin\phi_{c+}\cos\theta_{c-}\sin\phi_{s+}\sin\theta_{s-}\\
 O_{OAF} &\sim\cos\phi_{c+}\cos\theta_{c-}\cos\phi_{s+}\cos\theta_{s-}+
 \sin\phi_{c+}\sin\theta_{c-}\sin\phi_{s+}\sin\theta_{s-}\\
 O_{\pi BDW} &\sim\cos\phi_{c+}\cos\theta_{c-}\sin\phi_{s+}\sin\theta_{s-}+
 \sin\phi_{c+}\sin\theta_{c-}\cos\phi_{s+}\cos\theta_{s-}\\
 O_{FDW} &\sim\cos\phi_{c+}\sin\theta_{c-}\sin\phi_{s+}\sin\theta_{s-}-
 \sin\phi_{c+}\cos\theta_{c-}\cos\phi_{s+}\cos\theta_{s-}\\
 O_{SCs} &\sim\exp\imath\theta_{c+}\cos\theta_{c-}\sin\phi_{s+}\sin\phi_{s-}-
 \imath\exp\imath\theta_{c+}\sin\theta_{c-}\cos\phi_{s+}\cos\phi_{s-}\\
 O_{SCd} &\sim\exp\imath\theta_{c+}\cos\theta_{c-}\cos\phi_{s+}\cos\phi_{s-}-
 \imath\exp\imath\theta_{c+}\sin\theta_{c-}\sin\phi_{s+}\sin\phi_{s-}
\end{split}
\end{equation}
\end{widetext}

It is also useful to consider these operators in the $B_{o/\pi}$ basis
For example, the SDW operator in the $o$ band $SDW(o)$ and the CDW operator in the $\pi$ band $CDW(\pi)$
are
\begin{equation}
\begin{split}
O_{SDW(o)} &\sim\exp\imath(\phi_{c+}+\phi_{c-})\sin(\phi_{s+}+\phi_{s-}) \\
O_{CDW(\pi)} &\sim\exp\imath(\phi_{c+}-\phi_{c-})\cos(\phi_{s+}-\phi_{s-})
\end{split}
\end{equation}

To determine the phases, we need to obtain the exponents that characterize
the spatial decay of the operators' correlations . Using a standard procedure to compute
the correlations with the quadratic Hamiltonian \cite{giamarchi_book_1d} we find
\begin{equation}
\begin{split}
\eta_{CDW} &=K_{2}+K_{1}+K_{4}+K_{3} \\
\eta_{OAF} &=P_{1}^{*2}K_{2}+Q_{1}^{*2}K_{1}+P_{1}^{*2}K_{1}^{-1}+Q_{1}^{*2}K_{2}^{-1}+
              P_{2}^{*2}K_{4}+\\
      &       Q_{2}^{*2}K_{3}+P_{2}^{*2}K_{3}^{-1}+Q_{2}^{*2}K_{4}^{-1} \\
\eta_{SCd} &= K_{2}+K_{1}+K_{4}^{-1}+K_{3}^{-1}
\end{split}
\end{equation}
The exponents of the OAF and $\pi CDW$ fluctuations are the same, so we need to
evaluate logarithmic corrections to the powerlaw decay to determine
the dominant ordering.

\section{Simplified system of RG equations} \label{eq:simpl}

When $\cot2\alpha\rightarrow 0$ and $\cot2\beta\rightarrow\infty$
one gets the following system of first order RG equations for
the couplings
\begin{widetext}
\begin{equation}
\begin{split}
\frac{dg_{1c}}{dl} &=g_{1c}[2-(P_{1}^{2}K_{2}+\frac{1}{2}K_{3}^{-1}+\frac{1}{2}K_{4}^{-1})]
\\
\frac{dg_{1a}}{dl} &=g_{1a}[2-(P_{1}^{2}(K_{2}+K_{1}^{-1}))]
\\
\frac{dg_{2c}}{dl} &=g_{2c}[2-(\frac{1}{2}K_{3}^{-1}+P_{1}^{2}K_{1}+\frac{1}{2}K_{4}^{-1})]
\\
\frac{dg_{1}}{dl} &=g_{1}(2-(K_{2}+K_{1}))
\\
\frac{dg_{2}}{dl} &=g_{2}(2-(K_{2}+K_{1}))
\\
\frac{dg_{\parallel c}}{dl} &=g_{\parallel c}
(2-(P_{1}^{2}K_{1}^{-1}+\frac{1}{2}K_{3}^{-1}+\frac{1}{2}K_{4}^{-1}))
\end{split}
\end{equation}
\end{widetext}
The zeroth order approximation to the above system is obtained
using the fact that $K_{4}$ ($K_{2}$) is much smaller (larger) than one.

For the $C2S1$ phase, the relevance of the
important coupling needs to be checked. This gives us only one
differential equation in this case (assuming that close to the
fixed point $|g_{1}|=|g_{2}|=g$)

\begin{equation}
\frac{dg}{dl}=g(2-(K_{2}+K_{1}))+ P_{1}Q_{1}(K_{2}-K_{1})g
\end{equation}

Taking into account the fact that $P_{1}Q_{1}<0$, that it keeps decreasing during the flow, and that the initial $K_{2}$ makes \emph{g} irrelevant, one finds that
a significant decrease of $K_{2}$ would be required in order to make \emph{g} relevant.

\bibliographystyle{prsty}
\bibliography{totphys,ladder3b}

\end{document}